\documentclass[graybox]{svmult}

\usepackage{mathptmx}       
\usepackage{helvet}         
\usepackage{courier}        
\usepackage{type1cm}        
\usepackage{makeidx}         
\usepackage{graphicx}        
\usepackage{multicol}        
\usepackage[bottom]{footmisc}
\usepackage{hyperref}        
\usepackage{soul}            
\usepackage{ulem}
\hypersetup{colorlinks=true,urlcolor=blue}
\usepackage[square,numbers]{natbib}
\makeindex             

\def\gsim{\lower.5ex\hbox{$\; \buildrel > \over \sim \;$}}
\def\lsim{\lower.5ex\hbox{$\; \buildrel < \over \sim \;$}}

\begin{document}
\title*{Orbits and background of gamma-ray space instruments}
\author{Vincent Tatischeff\thanks{corresponding author}, Pietro Ubertini, Tsunefumi Mizuno and Lorenzo Natalucci}
\institute{Vincent Tatischeff \at Universit\'e Paris-Saclay, CNRS/IN2P3, IJCLab, F-91405, Orsay, France, \\ \email{vincent.tatischeff@ijclab.in2p3.fr}
\and Pietro Ubertini \at IAPS/INAF, via Fosso del Cavaliere 100, I-00133 Rome, Italy, \\ \email{pietro.ubertini@inaf.it}
\and Tsunefumi Mizuno \at Hiroshima Astrophysical Science Center, Hiroshima University, 1-3-1 Kagamiyama, Higashi-Hiroshima, Hiroshima, 739-8526, Japan, \email{mizuno@astro.hiroshima-u.ac.jp}
\and Lorenzo Natalucci \at IAPS/INAF, via Fosso del Cavaliere 100, I-00133 Rome, Italy, \\ \email{lorenzo.natalucci@inaf.it}}
%
%
\maketitle

{\bf Abstract} 
 Gamma-ray telescopes in space are bombarded by large fluxes of charged particles, photons and secondary neutrons. These particles and radiation pose a threat to the nominal operation of satellites and limit the detection sensitivity of gamma-ray instruments. The background noise generated in gamma-ray space detectors by impinging particles is always much higher than the astrophysical signal to be detected. In this chapter, we present the different types of orbits suitable for gamma-ray missions, discussing their advantages and disadvantages, as well as the value of experiments embarked in stratospheric balloons. We then review the physical properties of all the background components in the different orbits and the stratosphere. 
 
\vspace{0.3cm}
{\bf Keywords} 
Gamma rays; High-energy astrophysics; Gamma-ray telescopes; Instrument background; Low Earth orbits; High-Earth orbits; Cosmic rays; Solar energetic particles; Earth albedo
 
\section{Introduction}

Due to the atmospheric absorption, gamma-ray experiments in the energy range from $\sim 100$~keV to a few tens of GeV must operate in space or in the high stratosphere, at an altitude higher than about 30 km. But while placed in low-Earth orbits (LEO), typically between 400 and 600~km, gamma-ray space missions need to avoid the Van Allen radiation belts filled with energetic charged particles \cite{vanallen59}. Similarly, high elliptical orbits around Earth have to limit as much as possible the time passed near perigee (the point of closest approach to Earth), when the spacecrafts pass through the electron and proton belts surrounding the planet. On the other hand, satellites orbiting outside the Van Allen belts, at distances greater than about 40~000 to 60~000~km from Earth, are not shielded by the geomagnetic field, which offers a natural protection against Galactic cosmic rays and solar energetic particles.     

The environment surrounding high-energy experiments in space has two main effects: (i) radiation damage due to highly energetic particles and (heavy) ions impinging within the detectors active volume and electronics, and (ii) degradation of the instrument sensitivity due to high counting rate if compared with the usually weak signals from cosmic sources to be studied. To give an order of magnitude, the integrated gamma-ray flux of the Crab nebula~--~the most intense (quasi-)stationary source of the gamma-ray sky~--~is $F_{\rm Crab}(\ge 1~{\rm MeV}) \approx 2 \times 10^{-3}$~ph~cm$^{-2}$~s$^{-1}$, which can be compared to the flux of Galactic cosmic-ray particles (mostly protons) in the near-Earth interplanetary medium: $F_{\rm GCR}(\ge 1~{\rm MeV}) \approx 2$~--~$5$~particles~cm$^{-2}$~s$^{-1}$ (depending on the strength of the solar modulation). Energetic particles not only induce a prompt instrumental background when they deposit energy in gamma-ray detectors, but they can also be the origin of a delayed background in soft gamma-ray telescopes due to the activation of spacecraft materials. And yet, gamma-ray astronomers are now designing space missions with the aim of reaching the milliCrab sensitivity in the MeV range, that is detecting tiny fluxes of cosmic gamma-ray photons in a background of surrounding energetic particles that can be a million times more numerous!

The first question to consider when preparing a new gamma-ray space mission is the orbit. The pros and cons of placing heavy gamma-ray detectors and spacecrafts in low- versus high-Earth orbits were already a matter of debate in the 80's. Historically, most of the satellites for high-energy astronomy were placed in LEO because of the lower power needed to reach flight altitude. But at the turn of the millennium, two European Space Agency (ESA) missions, the X-ray satellite \textit{XMM-Newton} (0.15~--~15~keV) and the gamma-ray observatory \textit{INTEGRAL} (20~keV~--~10~MeV), were launched to high elliptical orbits, with perigee and apogee at the start of the missions of $7000 \times 114~000$~km and $2000 \times 160~000$~km, respectively. Spacecrafts in such orbits spend most of the time outside the radiation belts, which implies a higher average background, though more stable and constant with the yearly orbital evolution. The choice of the best orbit in which to insert a gamma-ray satellite remains a difficult task, depending on multi-parametric issues, including scientific requirements, like the required sky coverage and duration of uninterrupted observations, operational constraints like the tracking station availability, and programmatic constraints like the cost and availability of the launcher and the launch site.  

In Section~\ref{sec:orbits}, we first present the different types of orbits available for gamma-ray space missions. We also devote a sub-Section to stratospheric balloon-borne experiments (Section~\ref{sec:balloon}), which have played a major role in the development of gamma-ray astronomy. In Section~\ref{sec:background}, we present in detail the different sources of background of gamma-ray satellites. We discuss the origin and overall properties of the background particles and radiation in Sections~\ref{sec:extragalactic} to \ref{sec:secondaries}, and the effects of the internally induced radioactivity in space in Section~\ref{sec:activation}. Our conclusions are finally given in Section~\ref{sec:conclusions}. 

\section{Orbits of gamma-ray space missions}
\label{sec:orbits}

The Earth's atmosphere is completely opaque to gamma-ray radiation at ground level. Cosmic photons of energy $\gsim 100$~keV penetrate the atmosphere with an efficiency $> 30$\% up to an altitude of $\sim 30$~km, i.e. in the upper stratosphere. These photons can then be detected with large area sensors, although the radiation arriving from cosmic sources provide a minuscule flux if compared with the induced detector background due to the high fluxes of charged particles. For these reasons soft-gamma-ray experiments in the energy range between $\sim 100$~keV and a few MeV are operated in space or, for experiments requiring short observing time, in the upper stratosphere. Higher energy gamma-rays, up to 100 GeV, are in principle also detectable at stratospheric altitudes, even if the long observing time needed to detect cosmic sources restricts the usefulness of stratospheric balloon experiments, which maximum floating time is limited to a few weeks in the best case. High-energy gamma-ray experiments are thus mainly conducted with satellites orbiting out of the atmosphere.

\subsection{Low-Earth orbits}
\label{sec:leo}

An LEO is defined as a geocentric orbit at an altitude between about 200 and 2000~km. However, most gamma-ray space instruments launched in LEO are placed between 400 and 600~km. At these altitudes, the orbital speed is $\sim 7.6$~km~s$^{-1}$ and the orbital period $\sim 95$~min. 

Historically, the first scientific satellites to study solar and cosmic X- and gamma-rays were injected in LEO: the \textit{Orbiting Solar Observatory} (OSO) series launched from 1962 to 1975, the three \textit{Small Astronomy Satellites} (SAS) launched in 1970, 1972 and 1975, the three \textit{High Energy Astronomy Observatory} (HEAO) missions in 1977, 1978 and 1979, the \textit{Solar Maximum Mission} (SMM) in 1980, the \textit{Compton Gamma-Ray Observatory} (CGRO) in 1991, BeppoSAX in 1996, among others. Gamma-ray missions still operative from an LEO comprise AGILE (\textit{Astro‐Rivelatore Gamma a Immagini Leggero}) launched in 2007 and \textit{Fermi} in 2008. More recently, also a substantial number of CubeSats, small low-cost satellites, have been placed in this family of orbits.

\begin{figure}[t]
\centering
\includegraphics[width=0.69\textwidth]{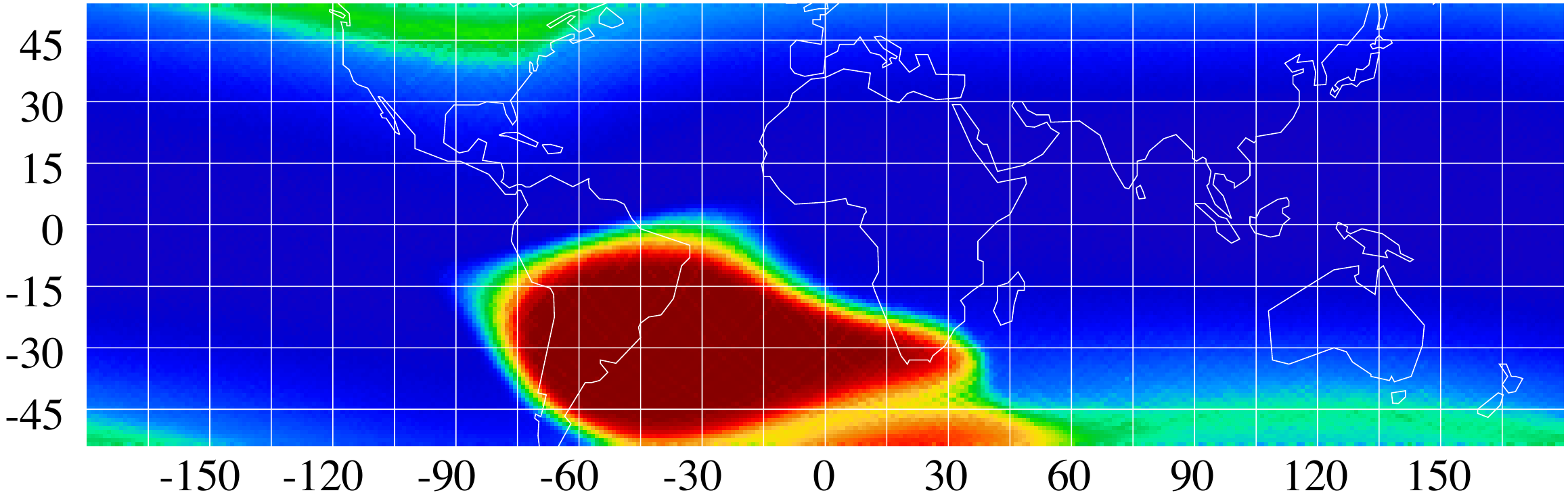}
\caption{South Atlantic Anomaly as detected by the particle background monitor aboard the \textit{ROSAT} X-ray Observatory, which operated for over eight years in the 1990's in a LEO of 580~km altitude and $53^\circ$ inclination (Credits: S.L. Snowden, ROSAT data).}
\label{fig:SAA1}
\end{figure}

One of the major advantage of LEOs is that the Earth's magnetic field works as a very effective shield against charged particles generated by the Sun and those arriving from the Galaxy. Unfortunately, the effect of the magnetic field shield is strongly modulated, not only in polar orbits (with an inclination of $60^\circ$--$90^\circ$ to the equator), but also in equatorial ones, which are usually chosen for high-energy instruments. The background count rate measured during the typical 95 minutes duration of the equatorial or low inclination orbits, is strongly modulated. This is basically due to the particular field line profile of the Earth's magnetic field funnelling charged particles in the south Atlantic region, the so-called South Atlantic Anomaly (SAA). As shown in Figure~\ref{fig:SAA1}, at an altitude of $\sim 580$~km, the SAA area spans a vast region extending from $-50^\circ$ to $0^\circ$ in geographic latitude and from $-90^\circ$ to $+40^\circ$ in longitude. 

\begin{figure}[ht]
\centering
\includegraphics[width=0.53\textwidth]{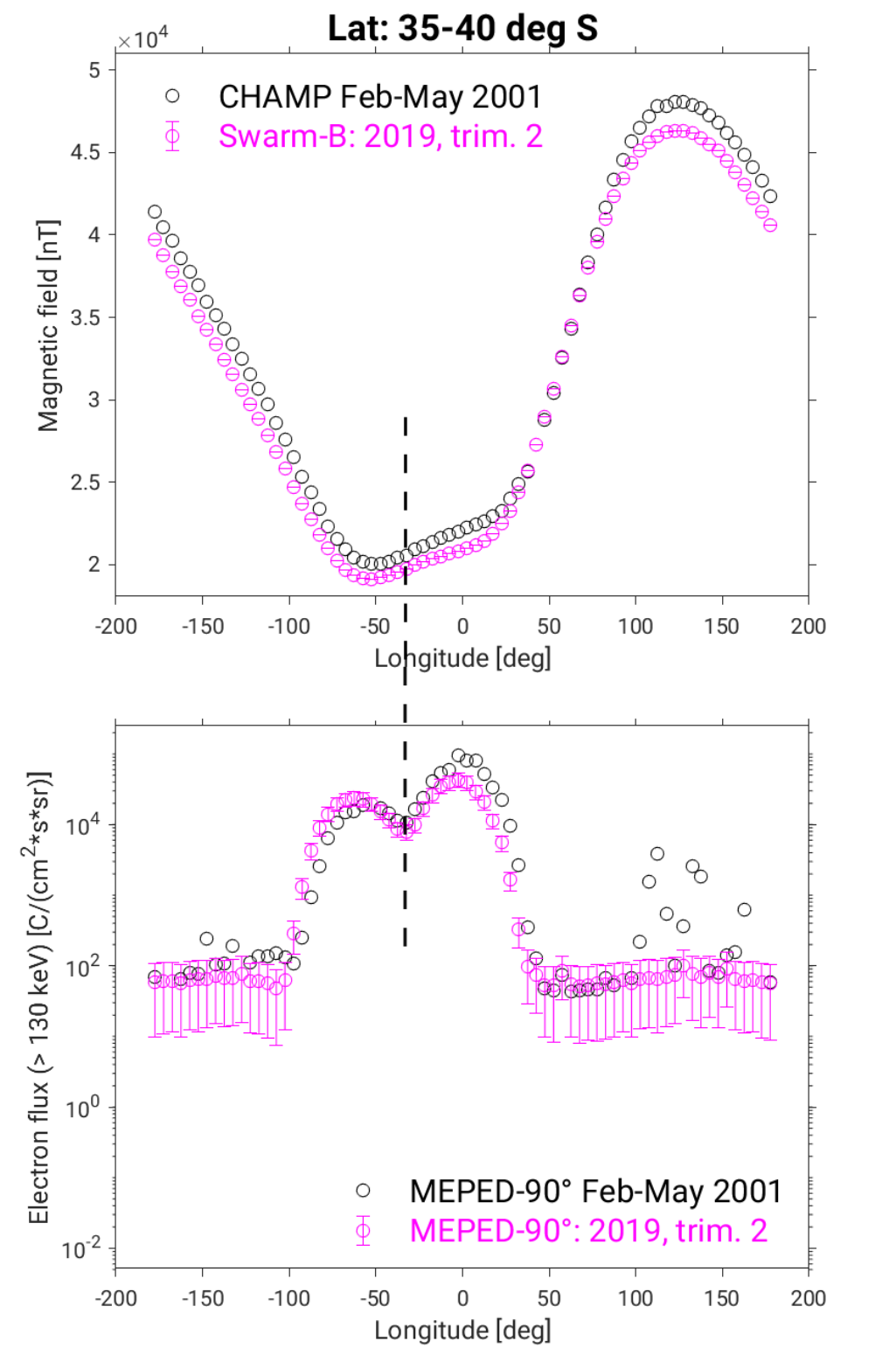}
\caption{\textit{Top}: Variation of the geomagnetic field at $\sim 800$~km altitude as a function of longitude in a latitude range of $35^\circ$~--~$40^\circ$ south crossing the core of the SAA (see Fig.~\ref{fig:SAA1}). \textit{Bottom}: corresponding integrated flux of electrons of energies $> 130$~keV as measured by the POES/MEPED-90$^\circ$ telescope anti-parallel to the satellite motion. All data were obtained in quiet condition, i.e. during periods of low solar activity. In each graph, the black symbols represent data collected in February~--~May 2001 (CHAMP satellite for the magnetic field and  NOAA16 for particles), while the magenta symbols show data collected in mid 2019 (Swarm-B for the B-field and NOAA19 for particles). Courtesy of A. Parmentier, private communication.}
\label{fig:SAA2}
\end{figure}

The SAA has been slowly expanding since the discovery of the radiation belts in 1958 (see \cite{vanallen59}). This is most likely related to the gradual weakening of the geomagnetic field, which has lost about $9$\% of its average strength over the last $200$~~years\footnote{See, e.g., \url{https://www.esa.int/Applications/Observing_the_Earth/FutureEO/Swarm/Swarm_probes_weakening_of_Earth_s_magnetic_field}}. The top panel of Figure~\ref{fig:SAA2} shows the longitudinal profiles of the geomagnetic field at $\sim 800$~km altitude and a latitude range deeply impinging the minimum of the SAA, as measured in 2001 with the CHAMP satellite and then in 2019 with the Swarm-B satellite. The bottom panel of this Figure shows the corresponding flux of energetic electrons of energies $> 130$~keV. We see that in just 18 years, a general lowering of the internal magnetic field is manifest, with major effects at the eastern SAA minimum. A corresponding variation in the relative intensities of electron flux peaks can be spotted, with cusp deepening (which suggests peaks moving away from each other) and global western drift of the entire SAA area. 

LEOs are particularly used for high-energy instruments given the relatively low background, despite the strong modulation over a time scale of tens of minutes. It should also be noted that the visibility of a satellite in an LEO by a ground station (such as the Malindi station operated by the Italian Space Agency) is limited to $\sim 10$~min per orbit, which requires an adequate on-board data storage and a fast download of the data at the passage over the receiving station. 

\subsection{High-Earth, highly elliptical and L1/L2 orbits}
\label{sec:ellipticalorbits}

A high-Earth orbit is a geocentric orbit at an altitude entirely above that of a geosynchronous orbit (35,786 km). To avoid the intense flux of energetic protons and electrons trapped in the outer Van Allen Radiation Belt, a spacecraft in such an orbit would have to be placed at least 10 Earth radii away ($\sim 64~000$~km). A more practical way to escape the Van Allen Radiation Belts for most of the orbit is to launch the satellite into a highly elliptical geocentric orbit of high eccentricity and high inclination. 

Several gamma-ray missions were launched to such high elliptical orbits, including COS-B in 1975, the first ESA mission to study cosmic gamma-ray sources, whose orbit perigee and apogee were about $350$~km and $100~000$~km, respectively.  COS-B spent about 80\% of its time outside the radiation belts, i.e. about 30 hours per 37.17-hour orbital period. Another example is the Soviet (later Russian) X- and gamma-ray space observatory GRANAT, which was placed in December 1989 in a highly eccentric 98-hour orbit with an initial perigee/apogee of $1760$~km/$202~480$~km, respectively, and an inclination of $51.9^\circ$. However, GRANAT's orbit has rapidly evolved due to lunar and solar perturbations: five years later, the perigee had increased to $59~000$~km and the inclination to $86.5^\circ$.

\begin{figure}[t]
\centering
\includegraphics[width=0.99\textwidth]{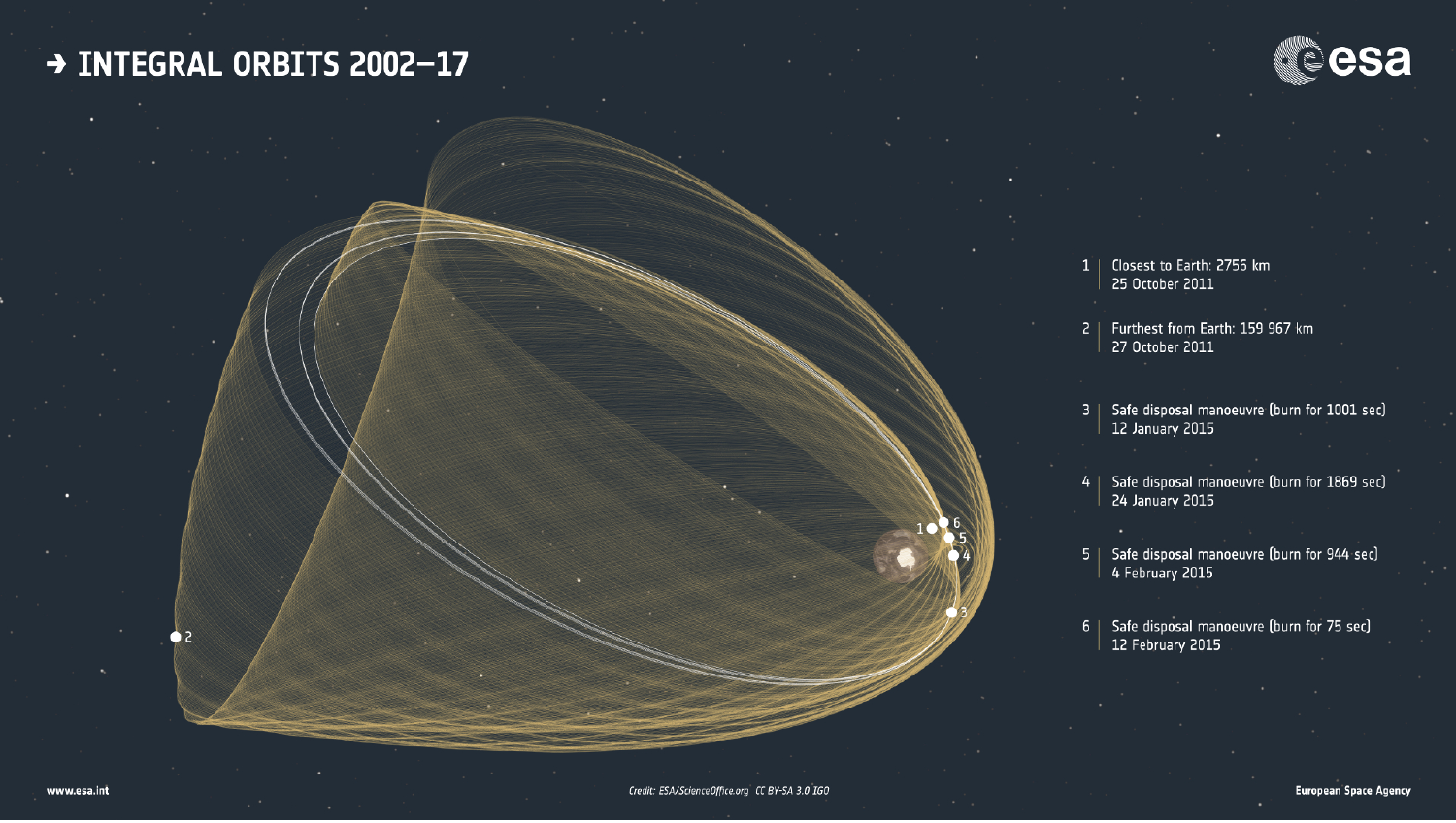}
\caption{\textit{INTEGRAL} orbit evolution after the launch, from the Baikonur Cosmodrome in Kazakhstan on 17 October 2002, up to October 2017. The spacecraft travels in a geosynchronous highly eccentric orbit with high perigee in order to provide long periods of uninterrupted observation with nearly constant background away from the radiation belts. Over time, the perigee, apogee and plane of the orbit evolve. So far, more than 19 years after injection in orbit, \textit{INTEGRAL} has travelled more than a billion km (Courtesy of ESA).}
\label{fig:integral_orbit}
\end{figure}

\begin{figure}[t]
\centering
\includegraphics[width=0.86\textwidth]{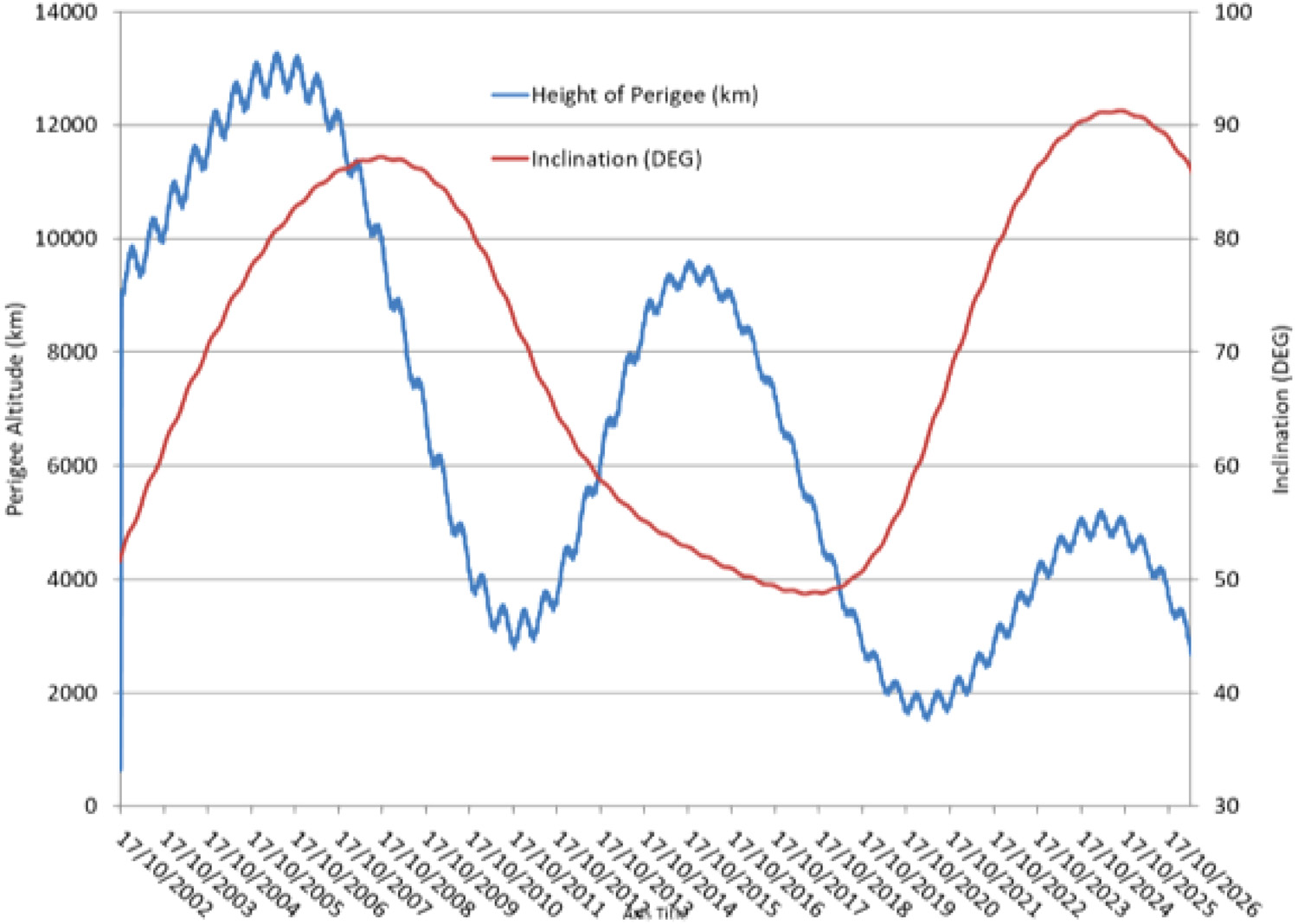}
\caption{Evolution of \textit{INTEGRAL}'s perigee altitude and orbital inclination (from Ref.~\cite{kuulkers21}).}
\label{fig:integral_orbit_2}
\end{figure}

The orbit of the \textit{INTEGRAL} observatory has also evolved over time since its launch in October 2002 (Figures~\ref{fig:integral_orbit} and \ref{fig:integral_orbit_2}). The choice of \textit{INTEGRAL}'s orbit before the launch was dictated by several arguments: i) orbit stability, without any limitation of the operational lifetime of the observatory, ii) long uninterrupted observations (about three days at the beginning of the mission), iii) very stable background, iv) low activation of the two high energy instruments SPI \cite{vedrenne03} and IBIS \cite{ubertini03}, v) possibility to have a continuous down and upper link with one main ground station, and other minor advantages \cite{winkler03}. With the chosen orbit the instruments can be operated above a distance from Earth of about 40~000~km, corresponding to $> 85$\% live observation time. As a result, on top of the almost constant background for the whole orbital duration (to date evolved to 2.8 days), the observatory is most of the time far from Earth with the possibility to observe the whole sky at the time. 
However, the predicted high background forced the two main instruments IBIS and SPI design to implement heavy active shielding for the high energy detectors, at the expense of sensitive detection areas.

After almost 20 years of scientific observations it is now clear that the orbital choice was key to the mission success. In fact, the nearly constant background (affected by almost negligible systematic errors), the exploitation of long duration uninterrupted observations, the very high fraction of observing time, and the fraction of time spent non-occulted by the Earth, have permitted it to achieve the most sensitive surveys of the Galactic centre and plane ever done \cite{bazzano06,bird16,malizia16,krivonos15,krivonos22} and to detect a wealth of gamma-ray bursts, including the first, and so far the only, gamma-ray counterpart of the merging of two neutron stars, i.e. the GW170817--GRB170817A event \cite{abbott17,savchenko17,ubertini19}. 

 Farther from Earth, the L1 and L2 Lagrangian points of the Sun-Earth system are attracting more and more interest for gamma-ray space missions. L1 and L2 are two points of gravitational equilibrium located at about 1.5 million km from Earth on the side and opposite side of the Sun, respectively. Among the scientific missions sent to L1 is NASA's \textit{Wind} spacecraft, comprising the Russian gamma-ray burst monitor Konus operating since 1994. More recently, in July 2019, the Russian–German soft and hard X-ray observatory Spektr-RG was launched in a halo orbit around L2. Although the background of gamma-ray instruments outside the Earth's radiation belts is higher than in LEO (see Section~\ref{sec:background} below), the L1 and L2 orbits have the advantage of greater stability in terms of illumination and thermal conditions. These orbits far away from Earth would be well suited to host a sensitive all-sky gamma-ray imager (with a field of view of almost $4\pi$~sr) for multi-messenger astronomy \cite{tatischeff19}.

\subsection{Stratospheric balloon experiments}
\label{sec:balloon}

\begin{figure}[t]
\centering
\includegraphics[width=0.66\textwidth]{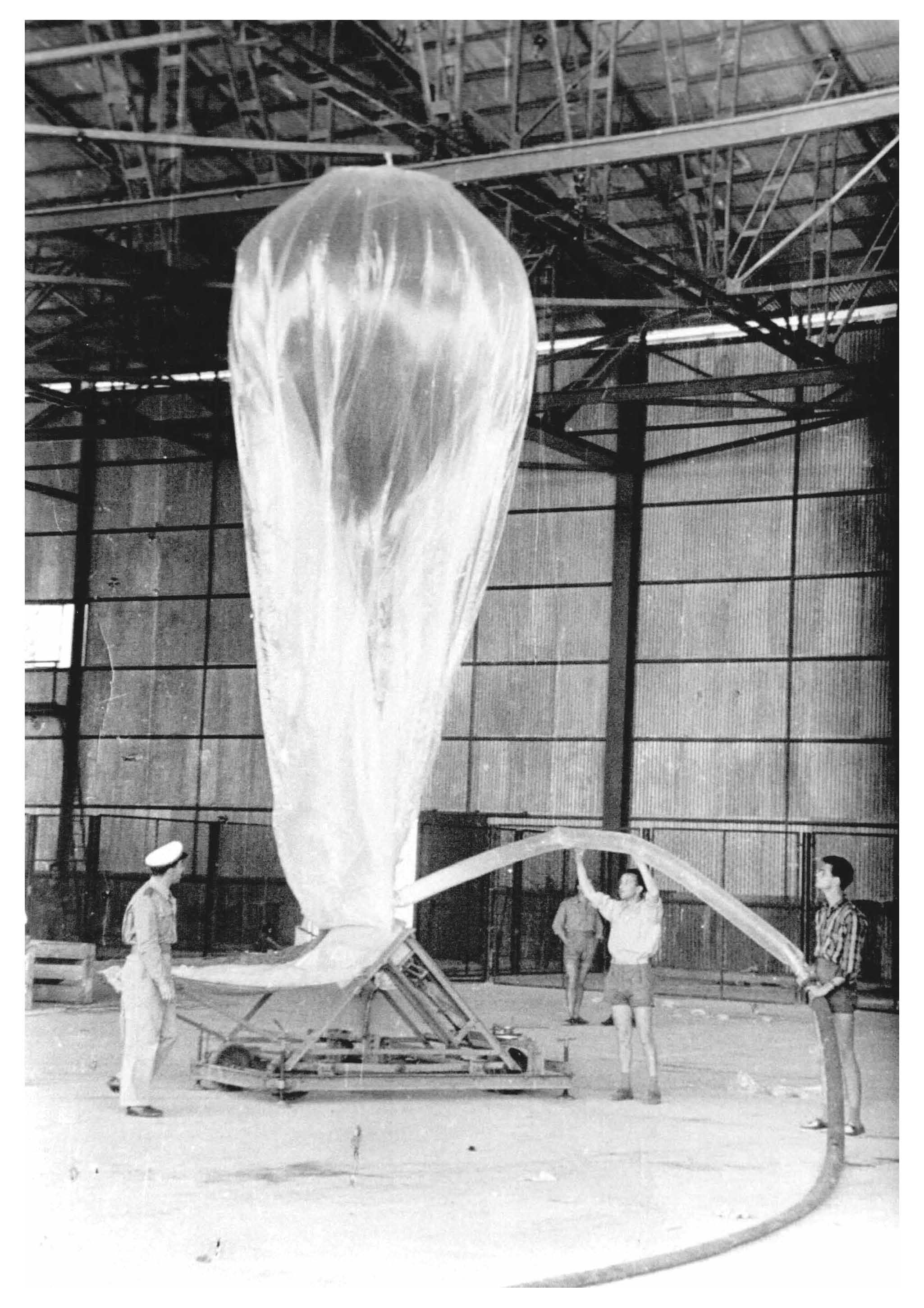}
\caption{The stratospheric balloon, with a volume of about 9~000 cubic meters, launched from the Cagliari Elmas Airport (Sardinia, Italy) in the early 1950s. Courtesy of P. Ubertini.}
\label{fig:cagliari}
\end{figure}

Stratospheric balloons, together with sounding rockets, have been the first historical attempt to understand if the ground measured radioactivity was of terrestrial or cosmic origin. In Figure~\ref{fig:cagliari} is shown one of the first balloon experiments, with a volume of 9~000 cubic meters, launched from Sardinia (Italy). The scientific objective of the experiment, designed and exploited under the lead of Prof. Edoardo Amaldi, was to study the production of  particles in the high atmosphere, by means of the use of emulsions. The flight was successfully exploited in 1953 (see \cite{ubertini08}).

The background rate of balloon borne instruments is influenced by the intensity of the incident cosmic-ray radiation, usually represented by the vertical cutoff rigidity, $R_\mathrm{cut}$, which is defined as the lowest rigidity required for a charged particle coming from the zenith direction in order to reach a given point on the Earth surface.  $R_\mathrm{cut}$ is a function of the geomagnetic latitude and varies with time, with secular changes that are on a time scale of a year or less. The vertical cutoff rigidity is calculated as a function of longitude and latitude through a standard model of the geomagnetic field, the International Geomagnetic Reference Field (IGRF) maintained by the International Association of Geomagnetism and Aeronomy (IAGA), which is updated regularly, the current model being the 13th generation IGRF \cite{alken21}. An approximate formula for estimating $R_\mathrm{cut}$ at satellite altitudes, assuming the Earth's magnetic field to be a simple dipole, is given in 
Section~\ref{sec:gcr} (Equation~\ref{eq:rcut}). 

\begin{figure}[t]
\centering
\includegraphics[width=0.95\textwidth]{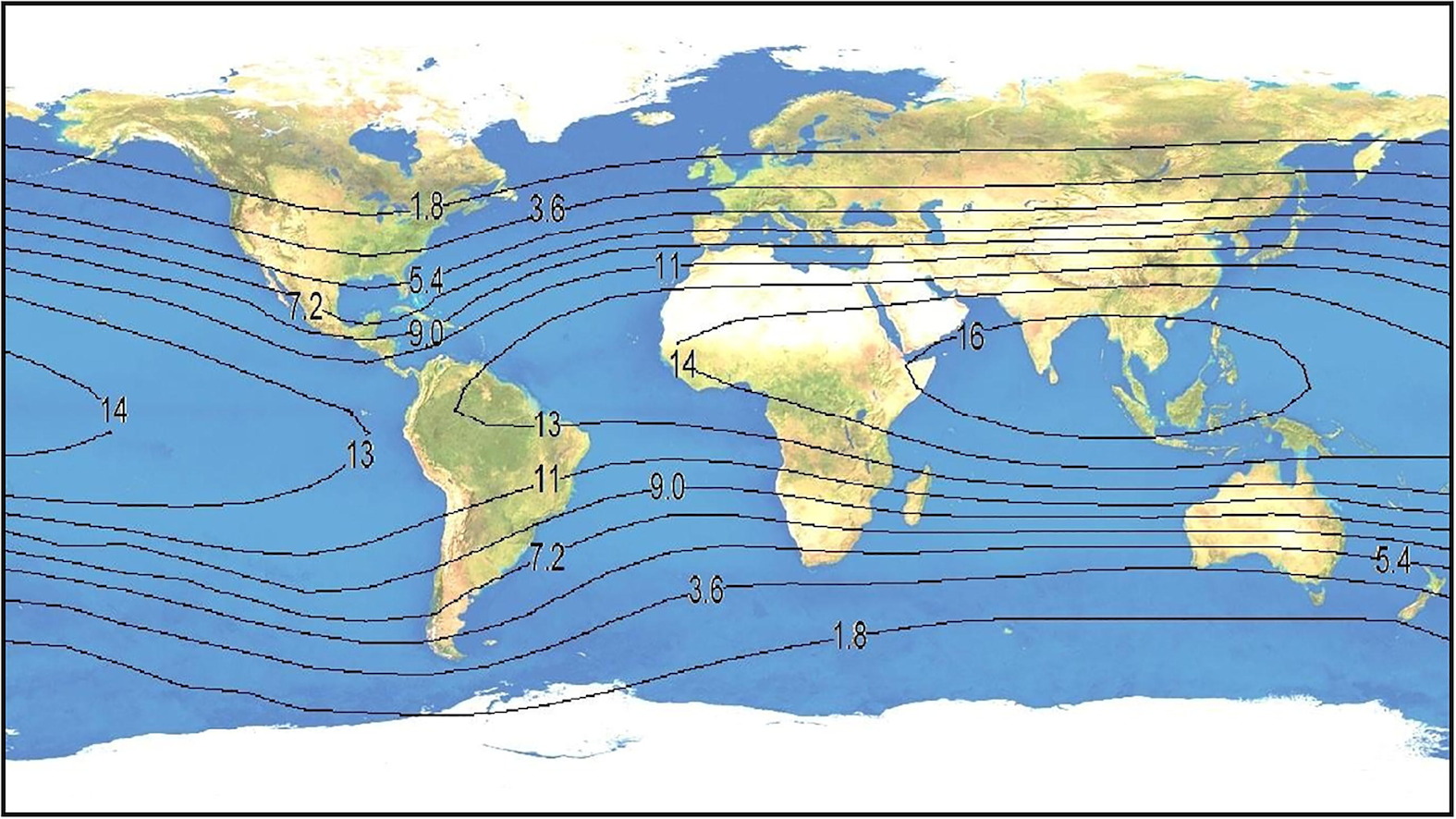}
\caption{Cutoff rigidity map for the year 2020 at the top of atmosphere (20~km altitude). Reproduced with permission from Ref.~\cite{gerontidou21}.}
\label{fig:rigidity_map}
\end{figure}

\begin{table}[t]
\caption{Vertical cutoff rigidity at main balloon launch sites}
\begin{tabular}{|c|c|c|c|c|} 
 \hline
 Launch site & Country & Latitude & Longitude & $R_\mathrm{cut}$ (GV)  \\ 
 \hline
  Kiruna & Sweden & 67.889 & 21.104 & 0.36 \\ 
 Aire-sur-L'Adour & France & 43.709 & -0.258  & 5.26 \\ 
 Timmins & Canada & 48.568 & -81.373 & 1.07\\ 
 CSBF Palestine & USA & 31.780 & -95.716 & 4.38 \\
 Fort Sumner & USA & 34.490 & -104.223 & 4.09 \\ 
 Alice Springs & Australia &  -23.799 & 133.883 & 8.06 \\ 
 Svalbard & Norway & 78.253 & 15.467 & 0.01 \\ 
 McMurdo & USA/Antarctica & -77.841 & 166.684 & 0.00 \\ 
 Syowa Station & Japan/Antarctica & -67.316 & 39.141 & 0.33 \\ 
 Hyderabad & India & 17.474 & 78.579 & 16.77 \\
 Taiki Aerospace & Japan & 42.500 & 143.433 & 7.78\\
 Wanaka Airport & New Zealand & -44.720 & 169.247 & 2.21 \\
 Spaceport Tucson & USA & 32.087 & -110.943 & 5.05\\  
 Trapani & Italy & 37.914 & 12.491 & 8.22 \\
 \hline
\end{tabular}
\\
\label{table:rigidity_at_sites}
\end{table}

\begin{figure}[t]
\centering
\includegraphics[width=0.85\textwidth]{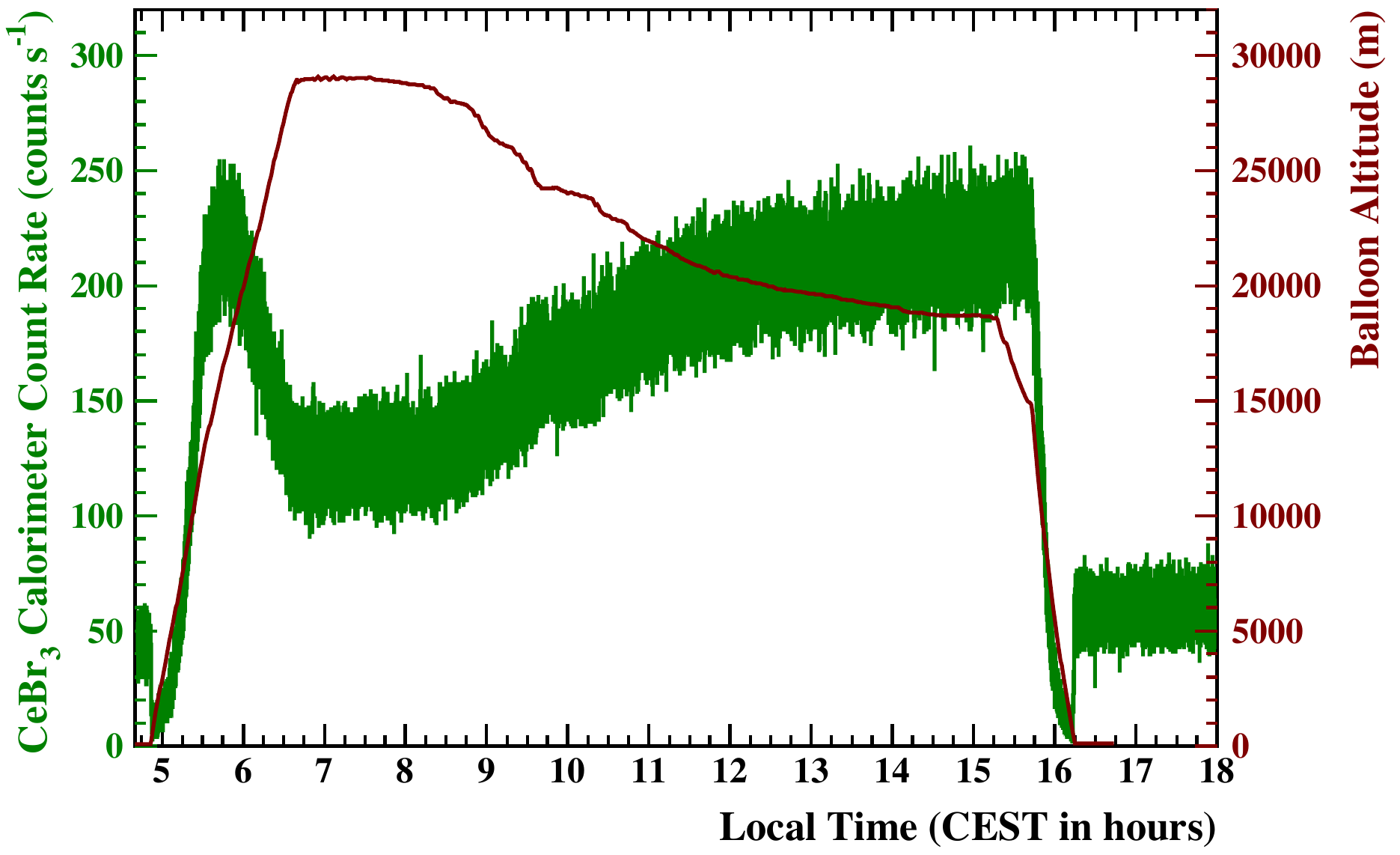}
\caption{Count rate in the energy range 40~--~1200~keV of a CeBr$_3$ scintillation detector (green curve) launched by a stratospheric balloon from the base of Aire-sur-L'Adour (France) and balloon altitude (red curve), as a function of local time during the flight on July 2, 2016. The maximum count rate is observed at an altitude between 15 and 20~km, i.e. at an atmospheric depth of $ \sim 100$~g~cm$^{-2}$, which corresponds to the Regener-Pfotzer maximum, i.e. the altitude at which the ionisation rate due to cosmic-ray interactions is the highest. At this altitude, the detector's count rate is about 5 times higher than that on the ground, which is mainly due to natural radioactivity. 
} 
\label{fig:balloon}
\end{figure}

A longitude~--~latitude map of cutoff rigidity is shown in Figure~\ref{fig:rigidity_map}. Values of $R_\mathrm{cut}$ span from nearly zero at the geomagnetic poles to $\sim 15$~GV at the magnetic equator. In Table~\ref{table:rigidity_at_sites}, we provide the cutoff rigidity values at the main balloon launch sites. Although the Earth's magnetic field shielding is less important at Antarctica and other near-pole stations (e.g. Kiruna, Svalbard, McMurod, Syowa Station), these launch sites can offer the advantage of long duration flights. In particular, ultra long duration balloons that could sustain flights for weeks/months are currently being developed by NASA (e.g. \cite{jones14J}). The duration of experiments launched from bases at lower latitudes is of the order of a few hours to a few days; as an illustration, the flight profile of an experiment launched in 2016 from L'Aire-sur-L'Adour (France) is shown in Figure~\ref{fig:balloon}.


Different cosmic-ray and gamma-ray experiments have been developed to be launched on stratospheric balloons, mainly in the framework of the NASA balloon program but also exploiting other launch opportunities. 
Recently, the EU has also created its scientific balloon infrastructure, HEMERA (e.g. \cite{raizonville19}). Balloon campaigns are a good opportunity to challenge the realisation of new experimental techniques in the gamma-ray domain. Relevant examples are the development of Compton and pair-production telescopes, high-resolution spectrometers based on cryogenic detectors, as well as gamma-ray polarimeters.

Since the 1960s, pioneering observations on balloons started to provide important results in different fields. In 1967, Fishman \& colleagues observed the Crab Pulsar in low-energy gamma rays yielding a measurement of the Crab spin period \cite{fishman69}, whilst Haymes, Johnson et al. detected emission with an energy of $\sim500$~keV from the Galactic Center in the early 1970s \cite{haymes75}. A series of balloon-borne measurements of the cosmic gamma-ray background as well as the atmospheric MeV background were performed in the 1970s (e.g., \cite{schoenfelder77,mandrou79}).

The GRIS instrument \cite{tueller88} developed by NASA/GSFC uses cooled Ge detectors surrounded by a thick NaI anticoincidence shields for high-resolution gamma-ray spectroscopy. It was flown two times in 1988 from Alice Springs providing measurements of gamma-ray lines from SN~1987A \cite{tueller90} and had at least seven other successful flights up to 1995. During these flights it also detected other nuclear lines like the positron annihilation line at 511~keV from the Galactic Center \cite{gehrels91}, and the $^{26}$Al line emission from the Galactic Plane \cite{teegarden91}. 

The Compton technique allows performing spectroscopy and imaging of the sky in the MeV regime as successfully implemented with the COMPTEL instrument \cite{schoenfelder93} on board CGRO. Further developments came from the Nuclear Compton Telescope (NCT, \cite{coburn05}) built by a collaboration led by UC Berkeley. It carries 12 crossed-strip cryogenic germanium detectors with 3-D position resolution. NCT was flown successfully in 2005 and 2009 from Fort Sumner, with an additional attempted flight from Alice Spring in 2010. The COSI experiment, its successor, carries newly designed high purity Ge Detectors (GeDs) as a pathfinder to the recently approved COSI-SMEX mission.

More recently, the Kyoto University developed a Compton telescope for soft gamma-rays in the framework of the SMILE project. The instrument consists of an Electron Tracking Compton Camera (ETCC), with a gaseous tracker and a position sensitive scintillation detector. The first prototype, SMILE-I was flown from the Sanriku Balloon Center of ISAS/JAXA in 2006, reporting observations of the diffuse cosmic background and atmospheric gamma-ray components \cite{takada11}. A recent flight of the prototype SMILE-2+ has occurred from Alice Springs in April 2018, reporting observations of the Crab Nebula and the Galactic Center (e.g. \cite{nakamura18}) and a further prototype, SMILE-3 is being developed for a long-duration balloon flight. 

At higher energies ($E>20$~MeV), the GLAST/BFEM balloon experiment \cite{thompson02} was flown in August 2001 from Palestine. It consisted of an engineering model of one tower element of the Large Area Telescope (LAT; \cite{atwood09}) currently flying on board the {\it Fermi} satellite. The instrument carried a complex combination of detector elements including a Si strip pair conversion tracker, a CsI calorimeter and an anticoincidence system. 

As a gamma-ray polarimeter, PoGOLite pathfinder \cite{pearce12} uses the Compton effect and subsequent photo-absorption to detect events within an array of 61 well-type phoswich detector cells made with plastic and BGO scintillators with BGO anticoincidence shield and polyethylene neutron shield. The instrument has an effective area for polarisation measurements of $\sim40$~cm$^2$ at $E = 40$~keV. POGOLite pathfinder was launched in 2013 from Esrange after two failed attempts, and reported a measurement of the Crab polarisation. More precise measurements were performed by a more advanced version, POGO+, launched in 2016 \cite{chauvin17}.  
  
Finally, one remarkable application in the domain of cosmic ray research is EUSO-Balloon, flying prototypes developed for the Extreme Universe Space Observatory onboard the Japanese Experiment Module (JEM-EUSO). Launched in 2014 from Timmins, Canada by the CNES balloon division, it provided measurements of the UV background in different atmospheric and ground conditions \cite{adams15}.

\section{Background components}
\label{sec:background}

We now discuss each background component, first presenting the origin and overall properties of the background particles and radiation, and then describing their spectrum and angular distribution (Sections~\ref{sec:extragalactic} to \ref{sec:secondaries}). In Section~\ref{sec:activation}, we discuss more specifically the delayed background in soft gamma-ray telescopes due to the activation of satellite materials.

\subsection{Extragalactic gamma-ray emission}
\label{sec:extragalactic}

The extragalactic gamma-ray emission is both a significant component of background for gamma-ray observations outside of the Galactic plane, as well as an important science topic, especially for astrophysics in the MeV gamma-ray range \cite{deangelis18}. The first evidence for an MeV diffuse emission emanating from beyond our Galaxy was found in 1974 with a balloon-borne Compton telescope \cite{schonfelder74}. At higher energies ($> 30$~MeV), 
an apparently isotropic emission of extragalactic origin was discovered by the \textit{SAS~2} satellite \cite{fichtel78},
and has been studied in detail by \textit{Fermi}-LAT \cite{abdo2010}. In the hard X-ray range ($< 100$~keV), the first reliable measurements of the cosmic background were performed in the mid-1970s with the \textit{HEAO-1} observatory, which had a dedicated masking mechanism to disentangle the cosmic emission from the instrumental background \cite{marshall80}. Without such a mechanism in recent space missions, a practical way of studying the diffuse emission is to use the Earth as a screen occulting part of the background sky. This technique was used in 2006 with the \textit{INTEGRAL} satellite thanks to four dedicated Earth occultation observations \cite{churazov07,turler10} (see also Frontera \& Santangelo in this Volume for the history of gamma-ray observations). 
 
Below $\sim 200$~keV, the cosmic background is generally understood to arise primarily from the 
integrated emission of unresolved active galactic nuclei such as Seyfert galaxies \cite[e.g.][]{gilli07}. Above $100$~MeV, blazars are thought to make an important contribution \cite[e.g.][]{ackermann15}. But many other sources can significantly contribute to the extragalactic gamma-ray emission, including radio galaxies, star-forming galaxies, thermonuclear supernovae and gamma-ray bursts (see Volume 3b of the Handbook for a review of all extragalactic sources). 

\begin{figure}[ht]
\centering
\includegraphics[width=0.9\textwidth]{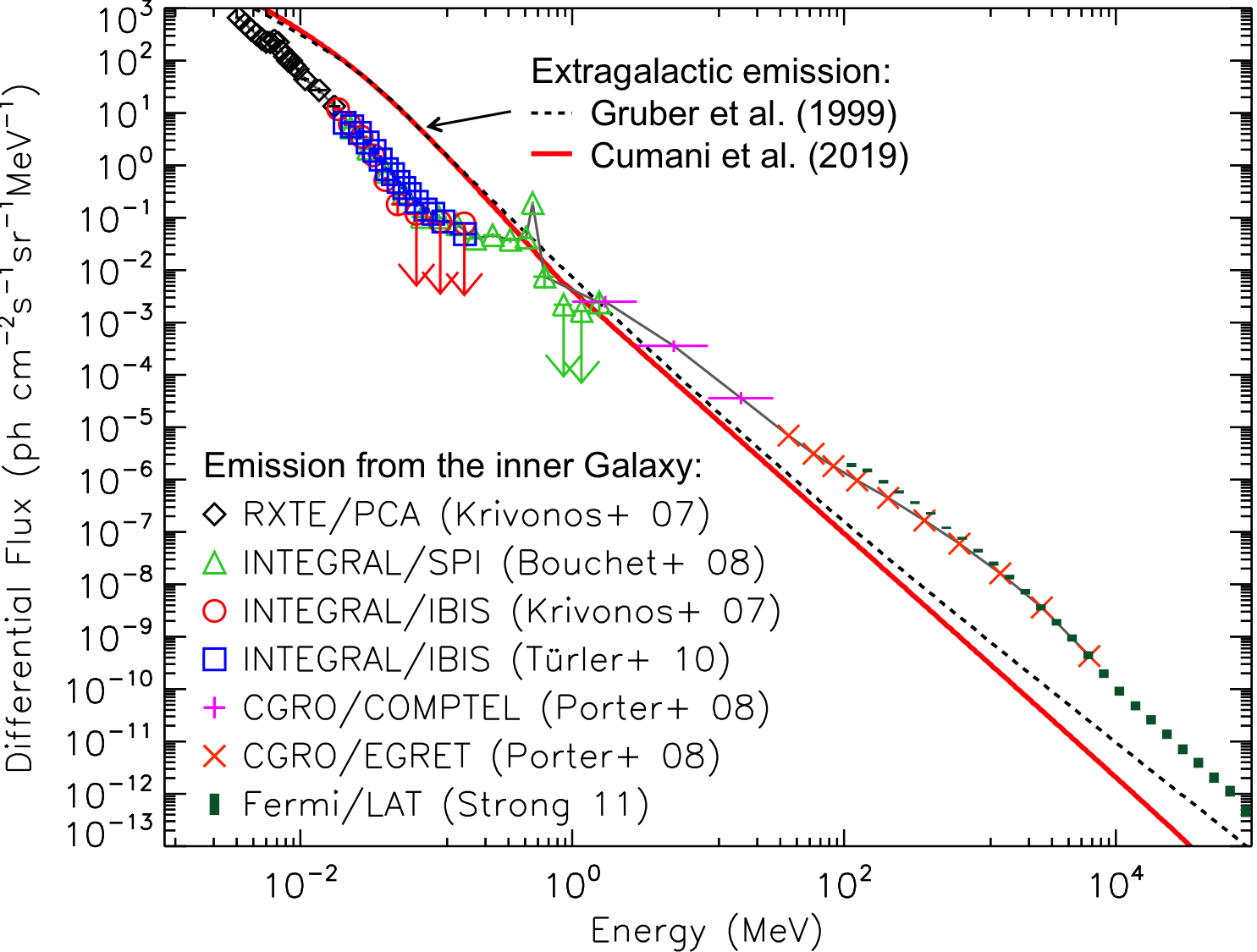}
\caption{X- and gamma-ray spectra of the emission from the inner Galaxy and of the isotropic extragalactic background. The Galactic emission data are taken from Galg\'{o}czi et al. \cite{galgoczi21} (see also Ref.~\cite{turler10}): \textit{RXTE}/PCA \cite{krivonos07}, \textit{INTEGRAL}/SPI \cite{bouchet08}, \textit{INTEGRAL}/IBIS \cite{krivonos07,turler10}, \textit{CGRO}/COMPTEL \cite{porter08} and \textit{CGRO}/EGRET \cite{porter08} for the region defined by $\mid \ell \mid < 30^\circ$ and $\mid b \mid < 15^\circ$, \textit{Fermi}/LAT \cite{strong11} for the region $\mid \ell \mid < 30^\circ$ and $\mid b \mid < 10^\circ$. Two models are compared for the extragalactic photon background: those of Gruber et al. \cite{gruber99} and Cumani et al. \cite{cumani19}. Adapted from Ref.~\cite{galgoczi21}.}
\label{fig:photon_bkg}
\end{figure}

The measured intensity of cosmic X- and gamma-rays was modeled with simple analytical functions by Gruber et al. \cite{gruber99} based on measurements by \textit{HEAO-1}, \textit{CGRO}/COMPTEL and \textit{CGRO}/EGRET instruments. This model is still often used to predict the sensitivity of planned space missions. In Figure~\ref{fig:photon_bkg}, it is compared to the more recent model given in Cumani et al. \cite{cumani19}, which is mainly based on \textit{INTEGRAL} and \textit{Fermi}/LAT observations. The two models agree to within 20\% below 300 keV, but deviate significantly at higher energies. The intensity modeled by Cumani et al. is lower than that of Gruber et al. by a factor of 1.8, 2.6 and 12 at 1~MeV, 1~GeV and 100~GeV, respectively. The model of Ref.~\cite{cumani19} should be preferred when studying the performance of future gamma-ray space missions.  

\subsection{Galactic gamma-ray emission}

The Galactic gamma-ray emission has been observed by many stratospheric balloon experiments and satellite missions since the 1960s (see Frontera \& Santangelo in this Volume). A compilation of data describing the emission from the inner Galaxy (in the Galactic latitude range $-30^\circ < \ell < 30^\circ$) is shown in Figure~\ref{fig:photon_bkg}. This emission mainly consists of unresolved sources from active stars, compact objects and supernova remnants, superimposed on a diffuse continuum radiation from cosmic rays interacting in the interstellar medium
(see \cite{churazov20,ackermann12} and Volume 3a of the Handbook for a review of all Galactic sources).
A large population of faint, unresolved sources is thought to make a major contribution to the diffuse-like emission below $\sim 100$~keV \cite{lebrun04}. The radiation from the central regions of the Galaxy also presents a strong line at 511~keV from positron annihilation (see the \textit{INTEGRAL}/SPI data in Figure~\ref{fig:photon_bkg}). 

As shown in Figure~\ref{fig:photon_bkg}, in the hard X-ray and soft gamma-ray domain the intensity of the Galactic emission is weaker than that of the extragalactic diffuse background. The integral flux above 10~keV of the emission from the inner Galactic disk ($\mid \ell \mid < 30^\circ$ and $\mid b \mid < 15^\circ$), which covers a solid angle of 0.542~sr, is 0.2~photon~cm$^{-2}$~s$^{-1}$ \cite{galgoczi21}. In comparison, the integral flux of the extragalactic background from the same region and the same energy range amounts to 2.0~photon~cm$^{-2}$~s$^{-1}$. The emission from the inner Galaxy is prominent at 511~keV and above about 1 MeV (Figure~\ref{fig:photon_bkg}). In the energy range covered by \textit{Fermi}/LAT ($E > 100$~MeV), it is more than an order of magnitude larger than the isotropic extragalactic background. In this energy range, the diffuse emission from the Milky Way can be the main source of background when observing a weak source in the Galactic plane. 

The high-energy gamma-ray emission from the most central part of the Galaxy is even brighter than that from the Galactic plane: the photon intensity detected by \textit{Fermi}/LAT from the Galactic center ($\mid \ell \mid < 2.5^\circ$ and $\mid b \mid < 1^\circ$) is higher than that from the Galactic disk ($\mid \ell \mid < 90^\circ$ and $\mid b \mid < 2^\circ$ excluding the region of $\mid \ell \mid < 2.5^\circ$ around the Galactic center) by a factor of $\sim 2.4$ between 100~MeV and 1~GeV, increasing to 3.4 at 100~GeV \cite{cumani19}.  

\subsection{Galactic cosmic rays and anomalous cosmic rays}
\label{sec:gcr}

Cosmic rays are composed of bare nuclei for about 99\% and electrons for the remaining 1\%. Among the nuclei, about 90\% are protons and 9\% are alpha particles. Most cosmic rays are produced in our Galaxy, probably in strong shock waves induced by supersonic stellar winds and supernova explosions \cite{tatischeff21}. Only ultra-high energy cosmic rays above $\sim 10^{17}$~eV are thought to be predominantly of extragalactic origin. Solar energetic particles emitted by our star are sometimes called ``solar cosmic rays". These particles and their effects on gamma-ray instruments are discussed in Section~\ref{sec:sep}.

Cosmic rays can have several adverse effects on gamma space telescopes. First they can change the state of components of electronic integrated circuits, causing electronic noise and errors such as single-event upset and latch-up. Space electronics uses radiation-hardened components that are made resistant to damage and malfunction caused by ionizing radiation. Charged cosmic rays can also trigger gamma-ray detectors in space and thus generate bad data contributing to the background noise. This background can be strongly suppressed by shielding the gamma-ray instruments with an anticoincidence detector highly sensitive to charged particles, but less sensitive to gamma-ray photons (see, e.g., \cite{moiseev07}). Also, cosmic rays can produce radioactive nuclei by spallation reactions in the satellite materials. These radioactivities can induce a delayed background in gamma-ray instruments against which it is more difficult to fight (Section~\ref{sec:activation}). 

An irreducible background for gamma-ray telescopes in space comes from cosmic-ray interactions in the outermost parts of the instrument, in particular the light shield, thermal blanket, and micrometeroid shield above the anticoincidence detector. Inelastic nuclear collisions of cosmic rays with these materials can produce gamma-ray lines in the MeV range and neutral pions that decay into gamma rays in the primary operating energy range of high-energy instruments. This high-energy emission was the dominant background component of the COS-B gamma-ray telescope for parts of the sky away from the Galactic plane \cite{strong87}. For the later developed gamma-ray instruments \textit{CGRO}/EGRET, \textit{Fermi}/LAT and \textit{AGILE}, a special effort was made to minimise the exterior material to help deal with this irreducible background. 

\begin{figure}[ht]
\centering
\includegraphics[width=0.49\textwidth]{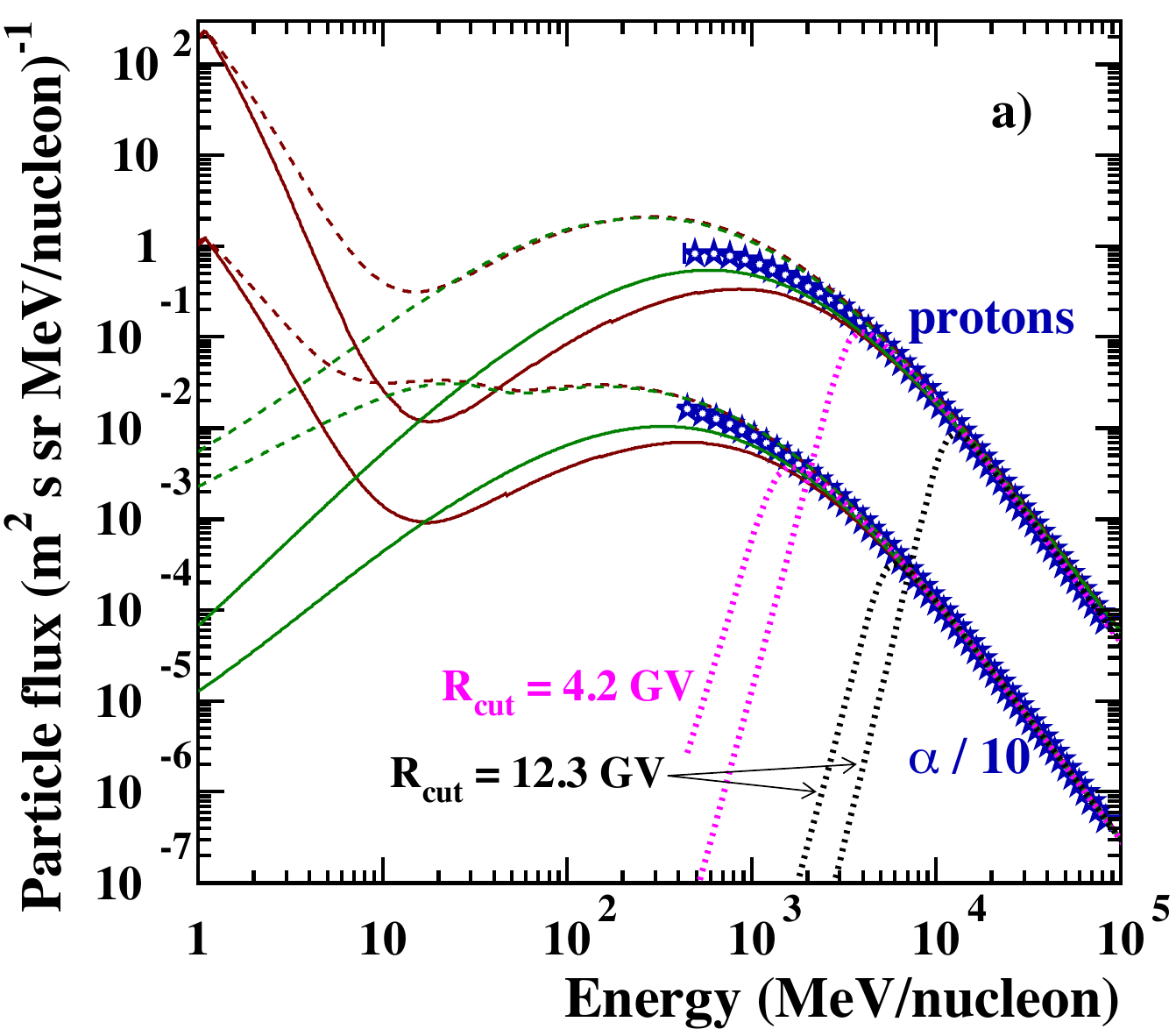}
\includegraphics[width=0.49\textwidth]{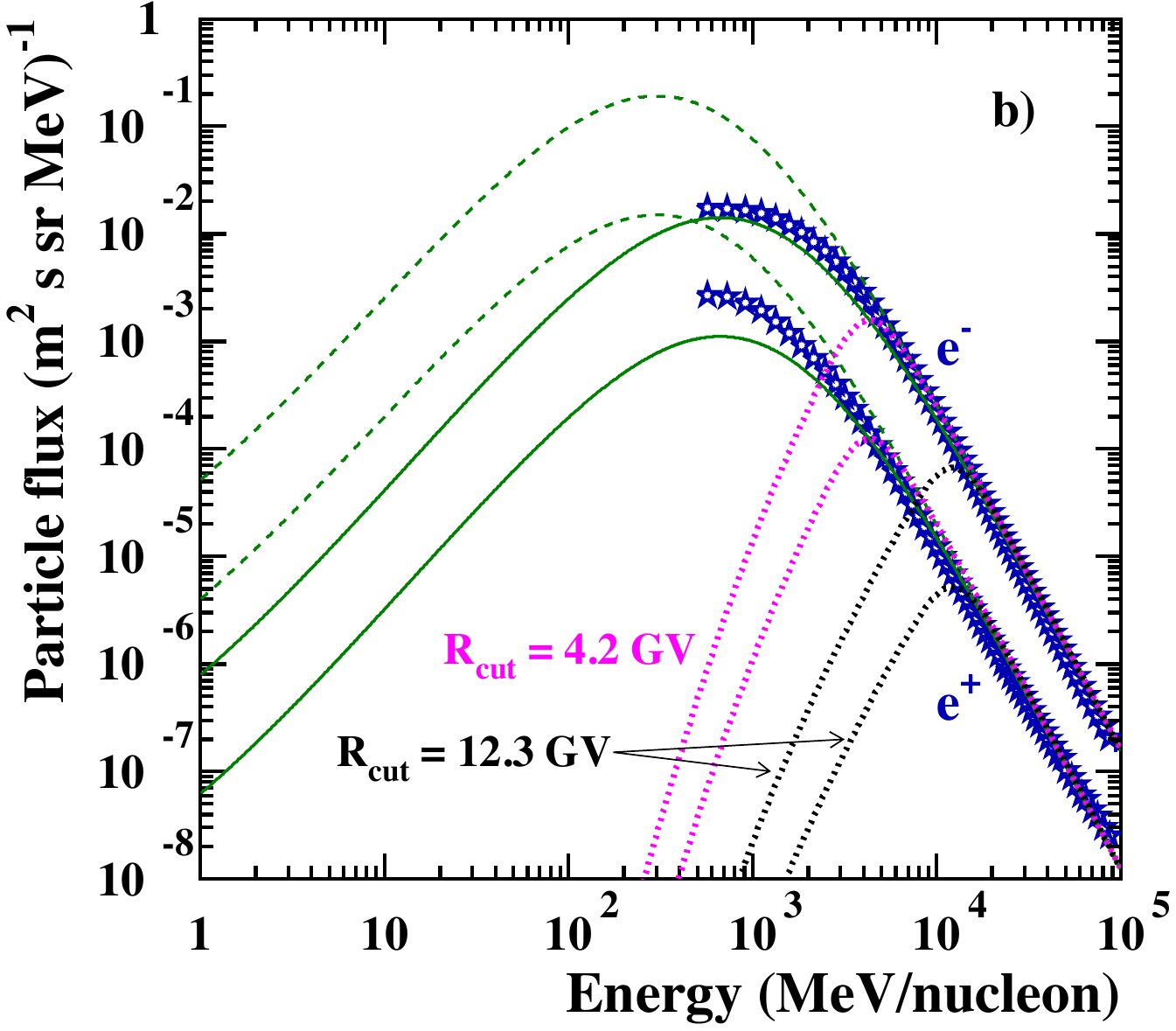}
\caption{\textit{(a)} Primary proton and $\alpha$-particle spectra in the near-Earth interplanetary medium (red and green lines) and in LEO for two geomagnetic cutoff rigidities: $R_{\rm cut}=4.2$~GV (magenta lines) and $R_{\rm cut}=12.3$~GV (black lines). Blue stars show the AMS-02 data from Refs.~\cite{aguilar15p,aguilar15a} ($\alpha$-particle intensities have been divided by 10 for clarity). The dashed and solid curves are for minimum and maximum solar activities, respectively. Green curves show the Galactic cosmic-ray model ISO-15390 used in CREME~2009 and red curves are from CREME96, which also includes anomalous cosmic rays (see text). \textit{(b)} Spectra of primary electrons and positrons. The AMS-02 data (blue stars) are from Refs.~\cite{aguilar19a,aguilar19b}. The electrons and positrons spectra in the near-Earth interplanetary medium (green lines) are calculated from Mizuno et al.~\cite{mizuno04}, with the solar modulation potential $\phi=500$ and $1100$~MV for minimum (dashed lines) and maximum (solid lines) solar activities, respectively.} 
\label{fig:gcr}
\end{figure}

Several dozen stratospheric balloon experiments and space instruments have been carried out to study cosmic rays, since the pioneering balloon experiments of Victor Hess in 1912. In the last decade, cosmic-ray flux measurements of extreme precision could be performed with the Alpha Magnetic Spectrometer (AMS-02), a state-of-the-art particle physics detector operating on the International Space Station since 2011. The spectra of cosmic-ray protons, alpha particles, electrons and positrons measured by AMS-02 \cite{aguilar15p,aguilar15a,aguilar19a,aguilar19b} are shown in Figure~\ref{fig:gcr}. 
 
The cosmic-ray flux changes with time in the interplanetary medium, because of the solar modulation effect caused by the magnetised solar wind against which cosmic rays must fight to reach the inner heliosphere. The cosmic-ray flux is found to be anticorrelated with the nearly periodic 11-year cycle of the solar magnetic activity. The force-field approximation \cite{gleeson68} provides a simple way to estimate the modulated cosmic-ray intensity in the near-Earth interplanetary medium, $F_{\rm mod}$, from the steady-state intensity of particles in the local interstellar medium (i.e. just outside the heliosphere), by assuming an effective shift of kinetic energy:
\begin{equation}
F_{\rm mod}(E)=F_{\rm LISM}(E+Ze\phi)\times\frac{(E+Mc^2)^2-(Mc^2)^2}{(E+Mc^2+Ze\phi)^2-(Mc^2)^2}~,
\label{eq:fmod}
\end{equation}
where $Ze$ and $M$ are the particle charge and mass, $c$ is the speed of light, and $\phi$ is a potential representing the level of solar modulation, which typically varies from $\sim 500$~MV for solar activity minimum to $\sim 1100$~MV for solar activity maximum. AMS-02 measurements of the cosmic-ray positron flux showed that $\phi$ reached a maximum in 2014 \cite{aguilar19b}, which corresponds to the peak of solar cycle 24.  

Satellites on LEOs are exposed to much lower fluxes of cosmic rays than spacecrafts outside the Van Allen Radiation belts (see Section~\ref{sec:orbits}), because of the additional modulation due to the Earth's magnetic field. The transmission of charged particles through the magnetosphere from the local interplanetary medium to a specific location defined by the satellite altitude $h$ and  the geomagnetic latitude $\theta_\mathrm{M}$ can be estimated from the geomagnetic cutoff rigidity \cite{smart05}:
\begin{equation}
R_\mathrm{cut} = 14.5 \times \left(1+\frac{h}{R_\mathrm{Earth}} \right)^{-2} \left(\cos \theta_\mathrm{M} \right)^{4}~\mathrm{GV}~,
\label{eq:rcut}
\end{equation}
where $R_\mathrm{Earth}=6371$~km is the Earth's radius. This equation takes the Earth's magnetic field to be a simple dipole. 
It should be modified for polar latitudes, where additional magnetospheric effects become important (see \cite{smart05}). The position of a satellite in the geomagnetic field is sometimes located using the McIlwain $L$-parameter (also noted $L_{\rm shell}$) rather than the geomagnetic latitude $\theta_\mathrm{M}$ (e.g. Figure~\ref{fig:2ndary_proton} below and Ref.~\cite{bidoli02}).


The suppression of particle flux at rigidities lower than $R_\mathrm{cut}$ can be modeled as:
\begin{equation}
F_\mathrm{cut}(E) = \frac{F_{\rm mod}(E)}{1+(R/R_\mathrm{cut})^{-r}}~,
\label{eq:fcut}
\end{equation}
where $R=pc/Ze$ ($p$ being the particle momentum) and $r=12$ for protons and alpha particles or $6$ for electrons and positrons \cite{mizuno04}. 

Intensities of protons, alpha particles, electrons and positrons in LEO are shown in Figure~\ref{fig:gcr} for two geomagnetic cutoff rigidities, $R_{\rm cut}=4.2$ and $12.3$~GV, which were obtained from Equation~\ref{eq:rcut} with $h=550$~km and $\theta_\mathrm{M}=40^\circ$ and $0^\circ$, respectively. In good approximation, the angular distribution of these particles can be assumed to be uniform in the range $0^\circ \leq \theta_\mathrm{z} \leq \theta_\mathrm{H}$ (and $0$ for $\theta_\mathrm{H} < \theta_\mathrm{z} < 180^\circ$), where $\theta_\mathrm{z}$ is the (telescope) zenith angle and
$\theta_\mathrm{H}$ is the horizon angle defined as the angle between the zenith and the top of the atmosphere ($H_\mathrm{A}=40$~km from sea level) at the satellite altitude $h$:
\begin{equation}
\theta_\mathrm{H}=90^\circ + \arccos\frac{R_\mathrm{Earth}+H_\mathrm{A}}{R_\mathrm{Earth}+h}~.
\label{eq:horizon}
\end{equation}

We see in Figure~\ref{fig:gcr} that the cosmic-ray intensities in LEOs are strongly suppressed below $\sim 1$~GeV, which can be a decisive asset for a gamma-ray mission. However, spacecrafts in LEOs are also exposed to secondary particles of different kinds, which are presented in Section~\ref{sec:secondaries}. But before this, we discuss various models for the intensities of primary particles in the heliosphere. 

\subsubsection{Protons and alpha particles}

In Figure~\ref{fig:gcr}a, we compare two models for the spectra of primary protons and alpha particles in the near-Earth interplanetary medium. The first one is the Galactic cosmic-ray model ISO-15390, which is the international standard\footnote{\url{https://www.iso.org/standard/37095.html}} for estimating the radiation impact of cosmic rays on hardware and other objects in space. We extracted this model from the online tool CREME~2009\footnote{\url{https://creme.isde.vanderbilt.edu/}}, which is the latest version of the cosmic-ray flux model in the CREME (Cosmic Ray Effects on Micro-Electronics) package frequently used in space dosimetry. We also used this tool to estimate the variation of the cosmic-ray intensities between the periods of minimum and maximum solar activity (see Figure~\ref{fig:gcr}). 

Figure~\ref{fig:gcr}a also shows primary proton and $\alpha$-particle spectra obtained from CREME96 \cite{tylka97}, which is the previous and still often used version of the CREME code. In addition to Galactic cosmic rays, this model also describes the intensities in the near-Earth environment of the so-called ``anomalous" cosmic rays, which are responsible for the strong rise of the total particle spectra with decreasing energy below about $10$--$20$~MeV~nucleon$^{-1}$ (red curves in Figure~\ref{fig:gcr}a). Anomalous cosmic rays were discovered with the IMP (Interplanetary Monitoring Platform) satellites in the early 1970s \cite{garcia73} and were soon interpreted as originating from interstellar neutral atoms that drift into the heliosphere, become ionized and are susequently accelerated somewhere in the outer heliosphere, perhaps in the solar wind termination shock \cite{fisk74}. They are generally considered to be unimportant for the assessment of radiation exposure in space and spacecraft design due to their low energies. However, they can constitute a significant source of background for hard X-ray and soft gamma-ray instruments in high-Earth orbits. 

\subsubsection{Electrons and positrons}

Energetic electrons and positrons are generally less important than ions for the design and operation of satellites due to their lower flux level and ionising power. However, they can significantly contribute to the background of gamma-ray instruments by emitting Bremsstrahlung radiation and 511~keV photons from interactions with satellite materials. 

We used the model of Ref.~\cite{mizuno04} to describe the intensity of primary electrons and positrons in the near-Earth interplanetary medium. Based on a compilation of measurements, this model takes the lepton spectra in the local interstellar medium to be power laws in rigidity: 
\begin{equation}
F_\mathrm{LISM}(E) = A_i \left(\frac{R}{\mathrm{GV}} \right)^{-3.3}~,
\label{eq:flisep}
\end{equation}
with $A_-=0.65$~particle~m$^{-2}$~s$^{-1}$~sr$^{-1}$~MeV$^{-1}$ for electrons and $A_+=0.051$~particle~m$^{-2}$~s$^{-1}$~sr$^{-1}$~MeV$^{-1}$ for positrons. The modulated and cutoff spectra shown in Figure~\ref{fig:gcr}b are then calculated from Equations~\ref{eq:fmod} and \ref{eq:fcut}.

\subsection{Solar energetic particles}
\label{sec:sep}

Solar energetic particles consist of protons, heavy ions and electrons with energy ranging from a few tens of keV to a few GeV. They are produced by solar flares in the low corona and by outward moving shock waves driven by Coronal Mass Ejections (CMEs) \cite{reames99}. The most harmful solar events for satellite operations and the background of gamma-ray instruments are caused by powerful interplanetary shocks driven by fast CMEs, which take about a day to arrive near Earth from their region of formation in the solar corona. A very active region of the Sun triggering several of such events can cause a strong disturbance of spacecrafts for more than a week (see, e.g., \cite{feynman00}). 

The strongest solar particle events produce a peak intensity of protons $\geq 10$~MeV near Earth $\gsim 4\times 10^8$~protons~m$^{-2}$~s$^{-1}$~sr$^{-1}$, as measured by the Geostationary Operational Environmental Satellites (GOES) since the 1970s\footnote{GOES 5-minute averaged integral proton fluxes are reported almost in real time at \url{https://www.swpc.noaa.gov/products/goes-proton-flux}}. In comparison, the intensity of Galactic cosmic-ray protons $\geq 10$~MeV at solar maximum is $1.3\times 10^3$~protons~m$^{-2}$~s$^{-1}$~sr$^{-1}$. Such extreme solar events are not expected to occur more than one or two times per solar cycle of 11 years (two such events occurred during solar cycle 22 in October 1989 and March 1991, but none since then). Relatively strong solar events producing more than $10^7$~protons~m$^{-2}$~s$^{-1}$~sr$^{-1}$ are more frequent: 13 during solar cycle 22, 17 during solar cycle 23, and 6 during solar cycle 24, which ended in 2019. These events can generate single-event upsets in on-board electronic systems and slightly degrade the efficiency of solar panels. Moreover, the data accumulated during these events are generally not usable for astronomy, because the background is too high. 

High-energy protons and heavy ions produced in strong solar particle events can occasionally enter the Earth's magnetosphere through the polar regions and interact with satellites in LEOs. Contrary to Galactic cosmic rays, solar energetic heavy ions are not fully ionised~--~they have a charge state characteristic of the $\sim 2\times10^6$~K coronal plasma from which they are accelerated \cite{leske95}~--~which increases their magnetic rigidity and thus enhances their ability to penetrate the Earth's magnetosphere (see Equation~\ref{eq:fcut}). Solar protons and heavy ions in the GeV range creating a nuclear cascade in the Earth's atmosphere are sometimes detected by neutron monitors as ``Ground Level Enhancements" above the background produced by Galactic cosmic rays. 

\subsection{Secondary particles in low-Earth orbits and the stratosphere}
\label{sec:secondaries}

Cosmic rays in or near the Earth's atmosphere consist of so-called primary and secondary components. The primary cosmic rays, mostly protons and alpha particles, are generated and propagate through extraterrestrial space (Section~\ref{sec:gcr}). When primary cosmic rays penetrate the air and interact with molecules, they produce low-energy particles (secondary cosmic rays). Hereafter we will describe the properties of each of particle species. We will first describe the origin and overall properties of the particle fluxes, then give their detailed properties based on measurements and simulations.

\subsubsection{Secondary protons}
Secondary protons are generated through the interaction of primary protons and alpha-particles with the Earth's atmosphere and are seen below the geomagnetic cutoff rigidity $R_\mathrm{cut}$ (Equation~\ref{eq:rcut}), with larger flux in higher geomagnetic latitude $\theta_\mathrm{M}$. They are trapped by Earth's magnetic field and have weak angular dependence.

\begin{figure}[ht]
\centering
\includegraphics[width=0.8\textwidth]{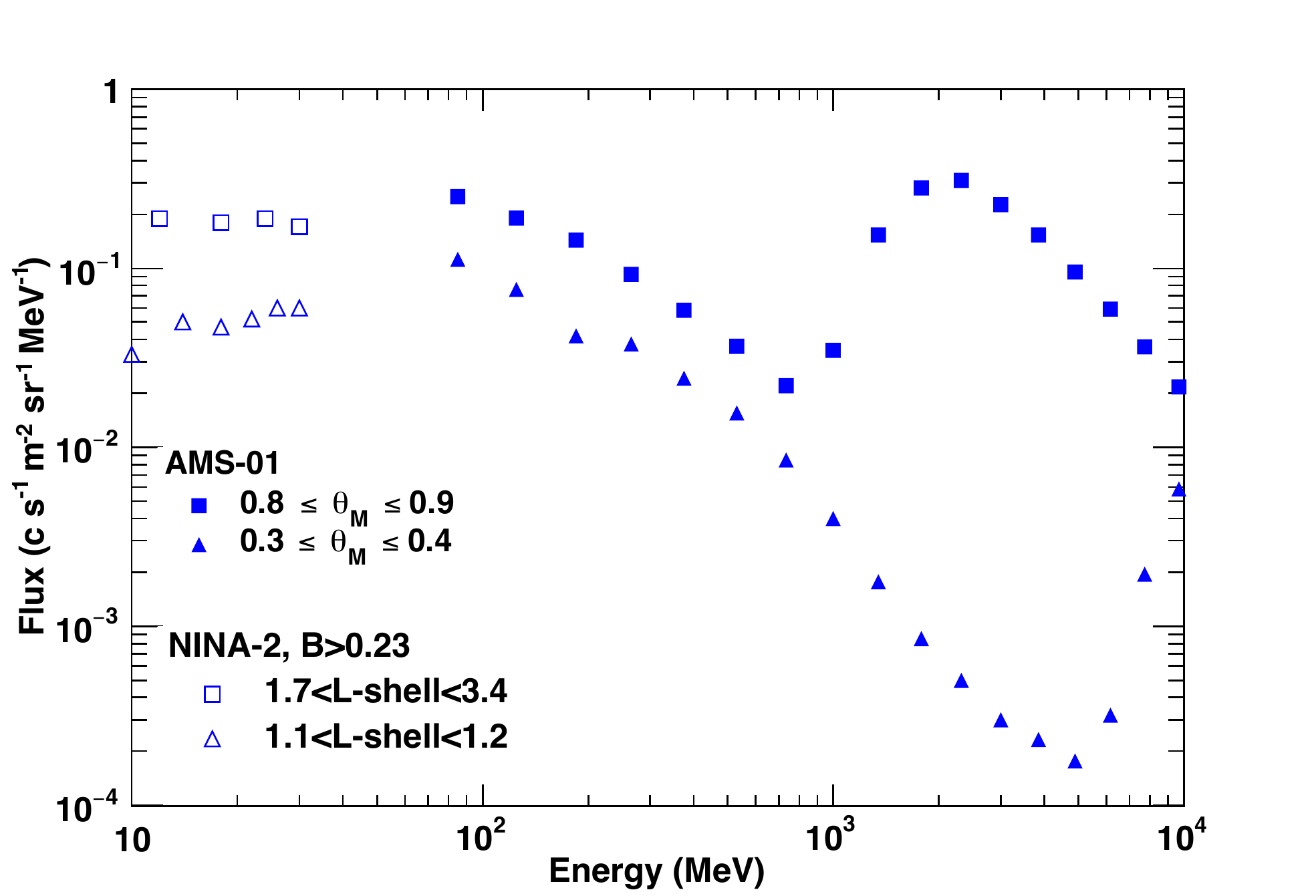}
\caption{Secondary (and primary) cosmic-ray proton spectra at LEO measured by AMS-01 and NINA-2 experiments. Vertically downward fluxes near the geomagnetic equator and pole are plotted.}
\label{fig:2ndary_proton}
\end{figure}

Vertically upward and downward spectra at LEO were measured precisely by the AMS-01 experiment at an altitude of {$\sim$}380~km for several $\theta_\mathrm{M}$ in the energy range {$E \ge$}100~MeV \cite{alcaraz00}. Below 100~MeV, they suffer from severe ionisation loss during propagation, and the spectrum flattens as measured by NINA/NINA-2 experiments \cite{bidoli02}. The spectra near the geomagnetic equator and pole are summarised in Figure \ref{fig:2ndary_proton}. 
Angular and altitude dependencies were studied by Zuccon et al. \cite{zuccon03} through a detailed simulation that takes account of the particle interaction and propagation in the Earth's atmosphere and magnetic field. They predicted small dependencies at altitudes above 200~km.
Secondary cosmic-ray protons ({$\le$}1~GeV) generated in the Earth's atmosphere have a small Larmor radius ({$\le$}100~km) and interact with the air. Hence their flux at balloon altitudes will be higher than that at LEO. Although the measured fluxes at balloon altitudes differ from experiment to experiment and therefore have large systematic uncertainty, they are much higher than that measured by AMS-01. See, e.g., Ref. \cite{mizuno04} and references therein
for more details. One can also use the 
Model for Atmospheric Ionizing Radiation Effects (MAIRE)\footnote{
\url{https://www.radmod.co.uk/maire}}
to calculate the altitude-dependent proton spectra 
(integrated over the upper and lower hemispheres)
in the atmosphere. Alternatively, one can use EXPACS\footnote{
\url{https://phits.jaea.go.jp/expacs/}}
\cite{sato2015} to calculate the spectrum as a function of $\theta_\mathrm{z}$ and the angle-integrated one.

\subsubsection{Secondary electrons and positrons}
Secondary electrons and positrons, produced in the Earth's atmosphere and trapped by the magnetic field, have similar but not identical properties to those of secondary protons. Firstly, they suffer from less ionisation loss and hence their flux near the geomagnetic equator (where Earth's magnetic field effectively trapped them) will be enhanced. Also, the electron flux is lower than the positron flux near the geomagnetic equator, since electrons oppositely suffer from the east-west effect for (parent) primary protons. Like secondary protons, angular and altitude dependence of secondary electrons/positrons are expected to be small. 

\begin{figure}[ht]
\centering
\includegraphics[width=0.49\textwidth]{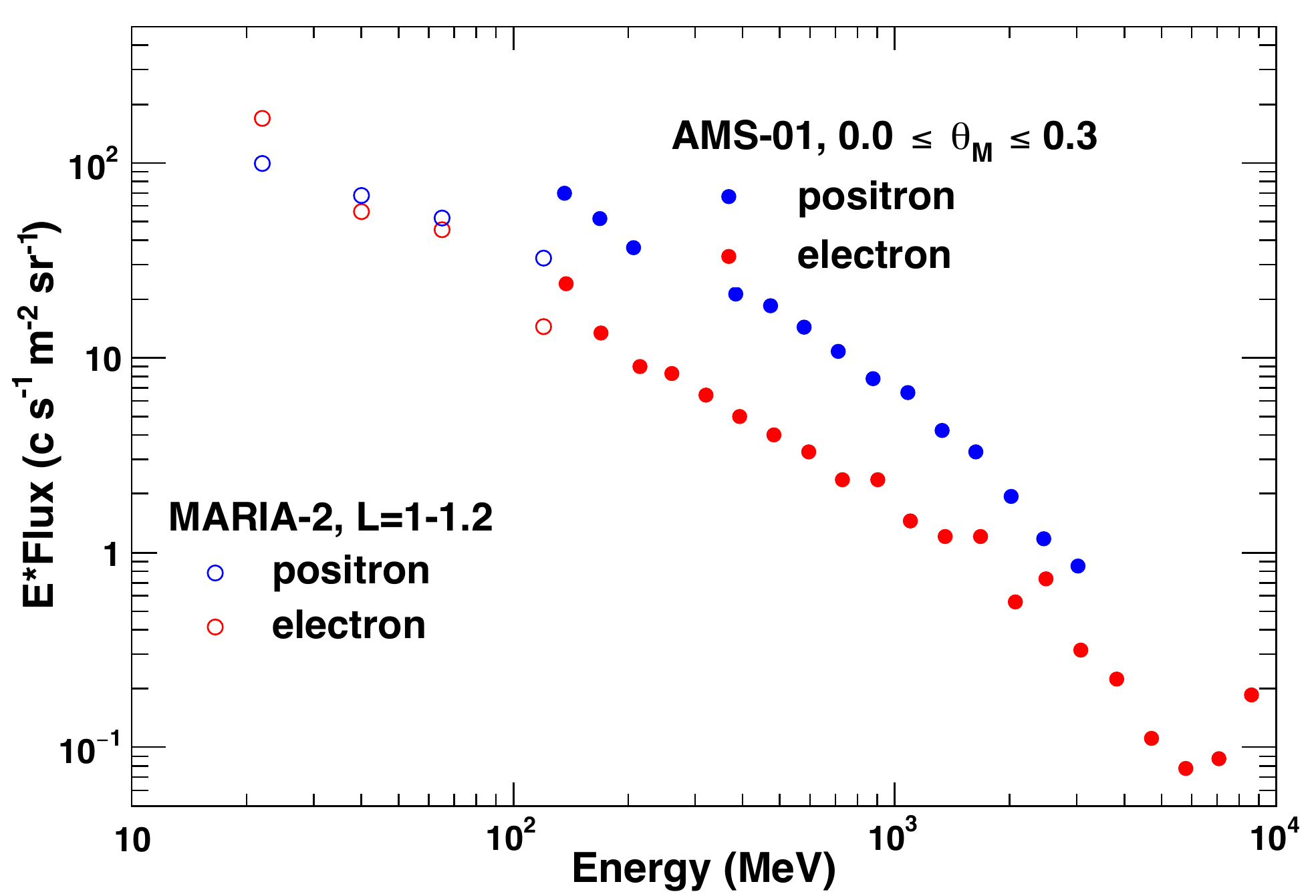}
\includegraphics[width=0.49\textwidth]{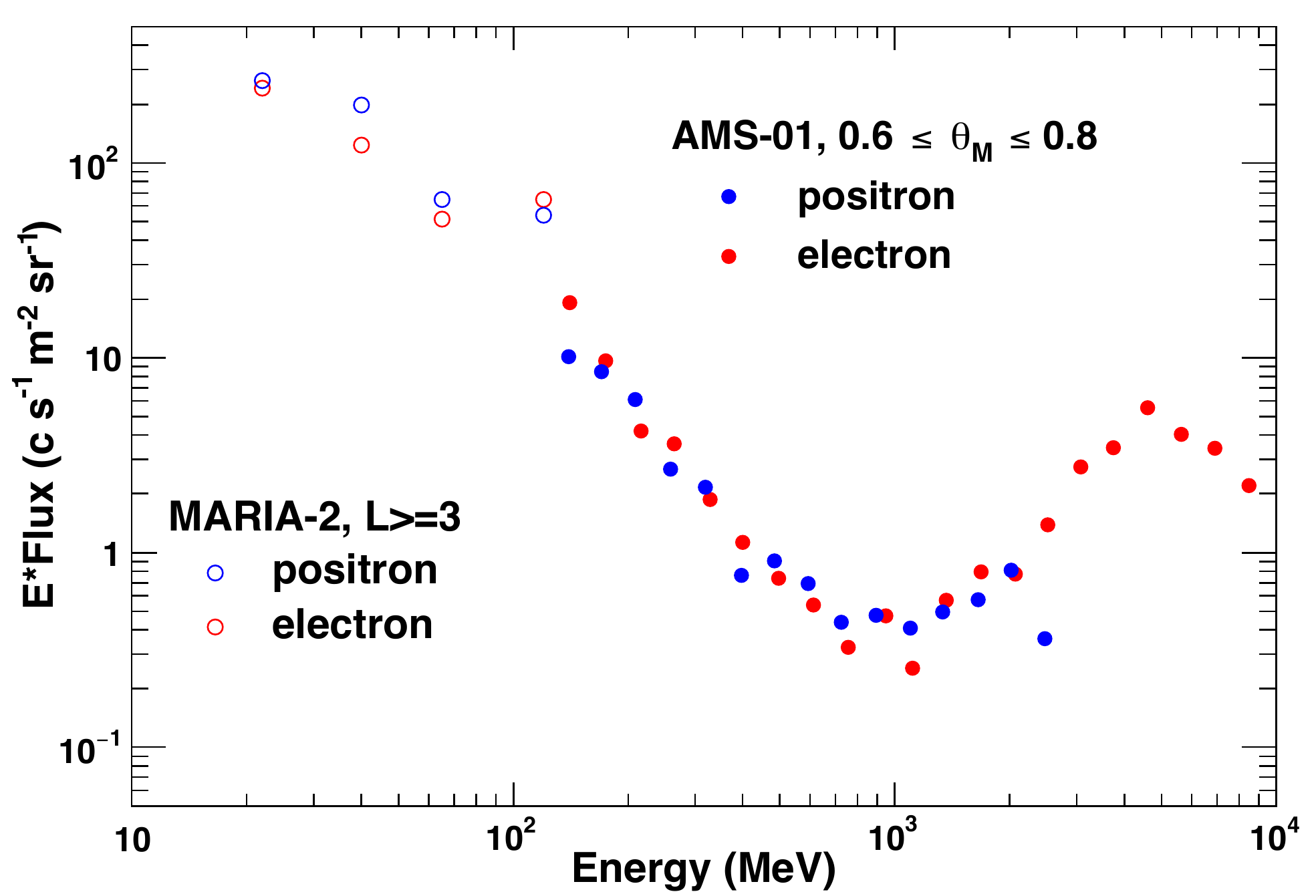}
\caption{Secondary (and primary) cosmic-ray lepton spectra at LEO measured by AMS-01 and MARIA-2 experiments. Vertically downward fluxes near the geomagnetic equator (left) and pole (right) are plotted.}
\label{fig:2ndary_lepton}
\end{figure}

Vertically upward and downward spectra were measured by the AMS-01 experiment at an altitude of {$\sim$}380~km for several $\theta_\mathrm{M}$ above 100~MeV \cite{alcaraz00b}. The spectrum flattens moderately below 100~MeV as measured by the MARIA-2 experiment \cite{mikhailov02}. The spectra near the geomagnetic equator and pole are summarised in Figure \ref{fig:2ndary_lepton}; the spectra at 100~MeV do not connect smoothly and hence have large uncertainty. Like for secondary cosmic-ray protons, the fluxes at balloon altitude are much higher than that measured at LEO by AMS-01 (see, e.g., \cite{mizuno04}). 
Again, one can also use 
MAIRE and EXPACS
to calculate the altitude-dependent lepton spectra in the atmosphere.

\subsubsection{Secondary gamma-rays (and X-rays)}

Secondary gamma rays ({$\ge$}100~keV) are produced either by interactions of cosmic-ray protons/alphas or bremsstrahlung of electrons/positrons in the atmosphere and hence have $R_\mathrm{cut}$ dependence similar to that of primary comic rays. Cosmic rays that enter the atmosphere near grazing incidence produce showers whose forward-moving gamma rays can penetrate the thin atmosphere, making the Earth's limb bright. The spectrum from the inner part of the Earth’s disk is softer since those gamma rays suffer from multiple scattering during the propagation in the atmosphere. Below about 100 keV (in X-rays), albedo of cosmic X-ray background (CXB) exceeds the atmospheric emission \cite{turler10}. Secondary X-rays and gamma-rays come from the direction below the horizon angle (i.e., $\theta_\mathrm{H} \le  \theta_\mathrm{z} < \pi$).

The spectrum and angular dependence of high-energy ({$\ge$}100~MeV) gamma rays at LEOs ($h \sim 550~\mathrm{km}$) were measured precisely by \textit{Fermi}-LAT \cite{abdo2009gamma}. It reported a sharp peak at the limb ($\theta_\mathrm{z} \sim 112^{\circ}$), and a power-law spectrum (above a few GeV) with a spectral index of about 2.8 (same as that of primary cosmic rays). The spectrum flattens below 1~GeV because of the cross-sections and kinematics of the hadronic interaction. It also reported a softer and weaker spectrum from the nadir direction. $R_\mathrm{cut}$ dependence was reported as $R_\mathrm{cut}^{-1.13}$ by \cite{gurian1979} above 80~MeV. The dependence is smaller at higher energies, as described by \cite{madlee2020}. Low-energy ({$\le$}100~MeV) gamma rays are produced by bremsstrahlung; they have a harder spectrum (e.g., \cite{imhof1976} \cite{ryan1977}) and a broader peak at the Earth’s limb (e.g., \cite{graser1977}). A similar $R_\mathrm{cut}$ dependence to that at a few 100s~MeV was reported by \cite{imhof1976}.

\begin{figure}[ht]
\centering
\includegraphics[width=0.8\textwidth]{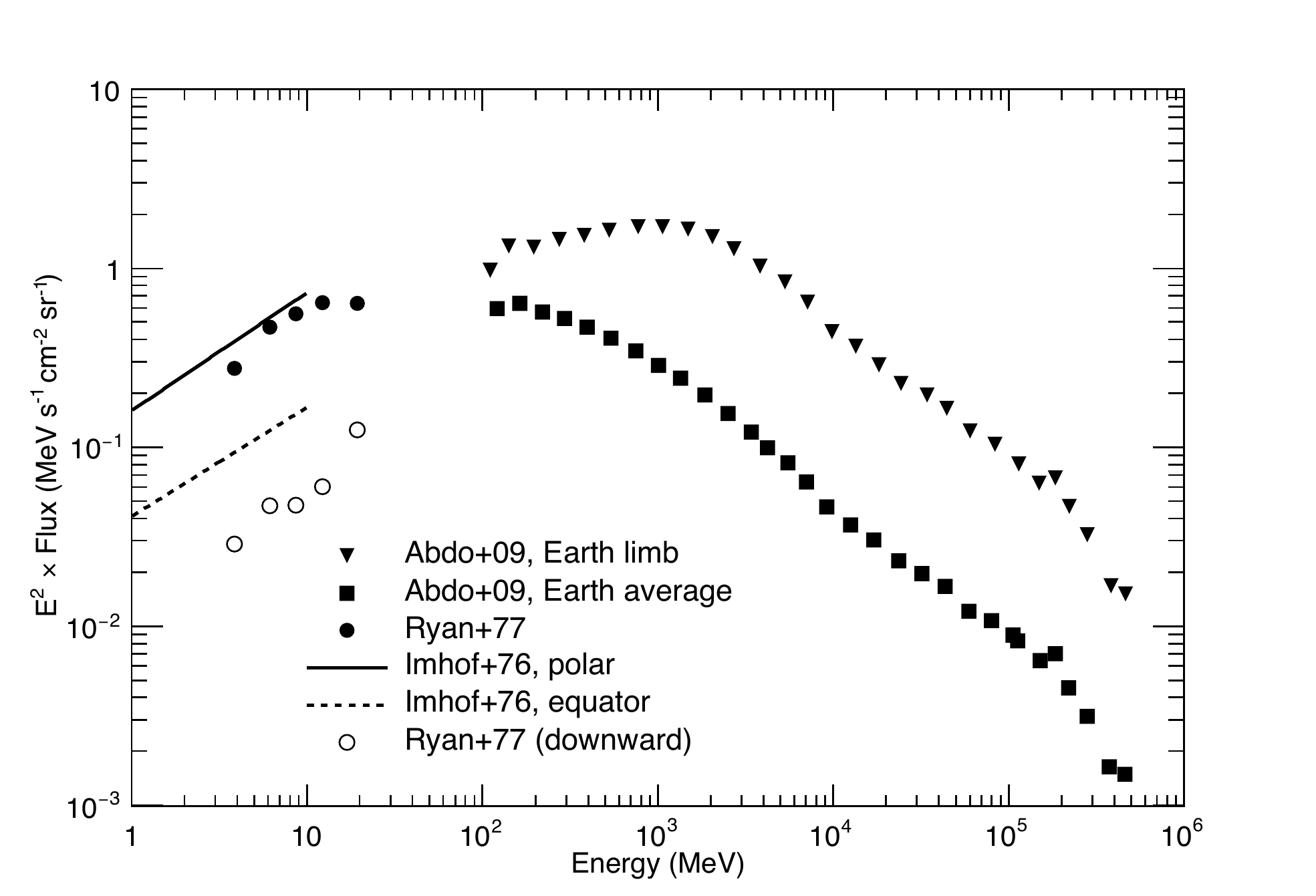}
\caption{Secondary upward gamma-ray spectra at LEO and balloon altitude. Abdo et al. (2009) \cite{abdo2009gamma} gives \textit{Fermi}-LAT measurements at the Earth's limb and the average over $\theta_\mathrm{z}=\mathrm{100-150~deg}$. 
Ryan et al. (1977) \cite{ryan1977} gives the result at Palestine, Texas ($R_\mathrm{cut}=\mathrm{4.5~GV}$). Imhof et al. (1976) \cite{imhof1976} gives the result by 1972-076B satellite below 2.7~MeV, and we extrapolated their best-fit formulae up to 10~MeV for reference. The downward spectrum by \cite{ryan1977} at $3~\mathrm{g~cm^{-2}}$ is also plotted for comparison.
}
\label{fig:2ndary_gamma}
\end{figure}

At balloon altitude, secondary upward gamma rays come from below the horizon ($\pi/2 \le \theta_\mathrm{z}$). Compared to the flux measured at LEO, those gamma rays have similar flux at the nadir direction and a less sharp peak toward the limb ($\theta_\mathrm{z}=\pi/2$). 
The downward gamma-ray flux is much smaller and is proportional to the atmospheric depth (e.g., \cite{daniel1974} \cite{schonfelder1980}). The gamma-ray spectra above 1~MeV at LEO and balloon altitude are summarised in Figure \ref{fig:2ndary_gamma}.
One may also use MAIRE and EXPACS to calculate the gamma ray spectrum in the atmosphere.

The CXB albedo was measured by the \textit{INTEGRAL} satellite. The flux is compatible with that of atmospheric emission at {$\sim$}50~keV (see Fig. 10 of \cite{churazov2007}). 

\subsubsection{Secondary neutrons}

Secondary neutrons are produced by cosmic-ray protons/alphas interactions in the atmosphere and have small chance of interaction during the propagation. Therefore, they have similar properties ($R_\mathrm{cut}$, angle, and altitude dependence) as those of secondary gamma rays. Although the neutron spectrum and direction are difficult to measure precisely, there are several observations as well as models that take into account the primary cosmic-ray spectrum and interactions/propagation in the atmosphere.

The model calculations by \cite{armstrong1973} and \cite{selesnick2007} give spectra below the top of the atmosphere; their spectra steepen progressively toward higher energies. The latter predicted a sharp peak at the Earth’s limb in {$\ge$}1~GeV. The limb emission is broader below that energy. Ref. \cite{morris1995} reported an $R_\mathrm{cut}$ dependence similar to gamma rays. At balloon altitude, there is also a downward flux proportional to the atmospheric depth (e.g., \cite{preszler1976}).

\begin{figure}
    \centering
\includegraphics[width=0.8\textwidth]{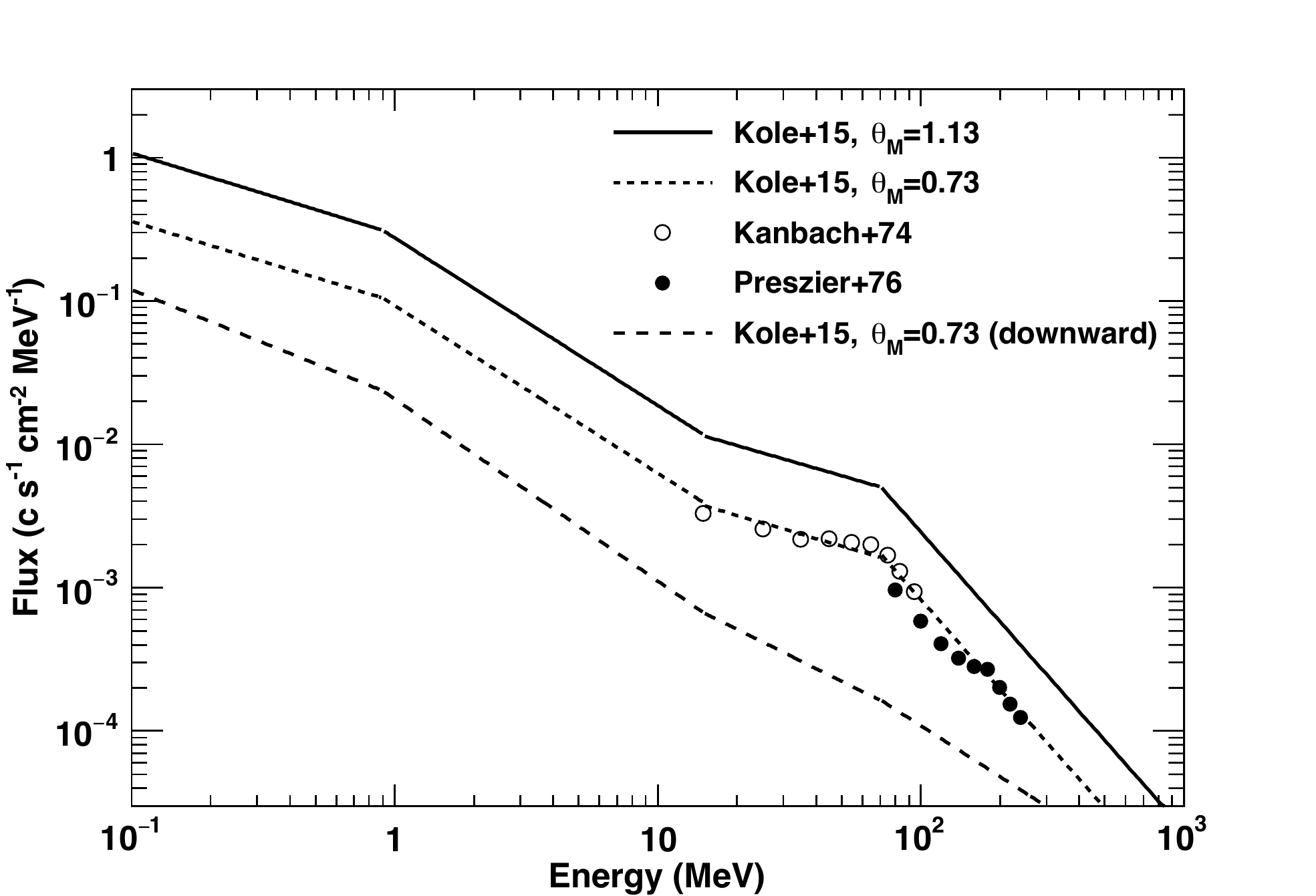}
    \caption{Secondary upward neutron spectral model by Kole et al. \cite{kole2015} (at $R_\mathrm{cut}=4.5~\mathrm{GV}$ and high latitude) and measurements at $R_\mathrm{cut}=4.5~\mathrm{GV}$. The downward spectral model at $4~\mathrm{g~cm^{-2}}$ is also plotted for reference.}
    \label{fig:2ndary_neutron}
\end{figure}

Ref. \cite{kole2015} employed a PLANETOCOSMICS package \cite{desorgher2006} to model the secondary neutron spectrum below the top of the atmosphere. They provided settings to calculate upward and downward fluxes integrated over $2\pi$ with $\cos \theta_\mathrm{z}$ multiplied (sometimes called as upward/downward current). As summarised in Figure \ref{fig:2ndary_neutron}, their upward spectral model agrees well with measurements at similar geomagnetic locations. One can use their model to predict the upward spectrum at LEOs with a smaller extent of zenith angle. The model spectrum of downward neutrons is also plotted for reference.
One may also use MAIRE and EXPACS to calculate the neutron spectrum in the atmosphere.
\\

\subsubsection{Particles trapped in the inner Van Allen radiation belt}
\label{sec:saa}

\begin{figure}[t]
\centering
\includegraphics[width=1.0\textwidth]{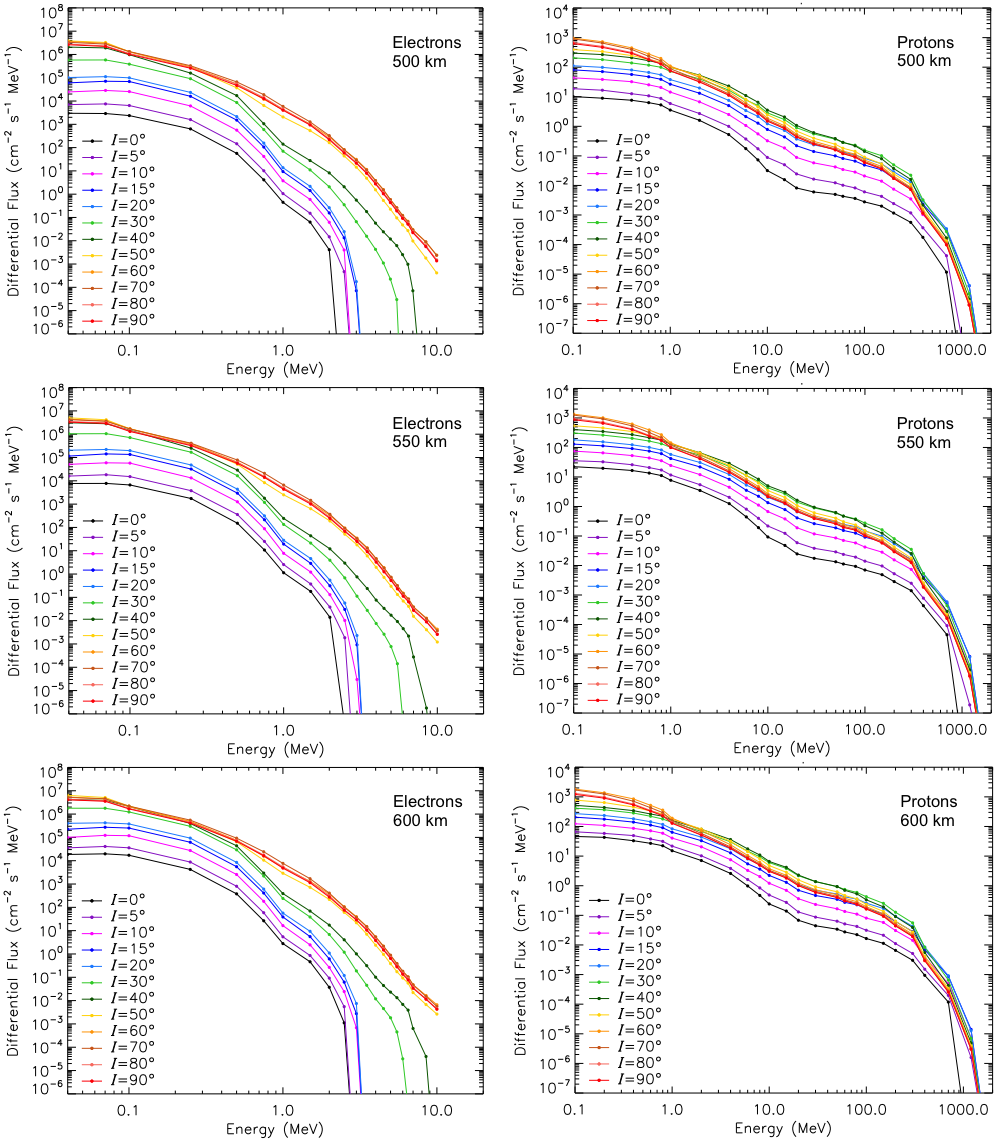}
\caption{Orbit-averaged differential fluxes of electrons (\textit{left panels}) and protons (\textit{right panels}) trapped in the inner radiation belt, calculated with the AE9 and AP9 models, respectively, for three different orbit altitudes (from \textit{top} to \textit{bottom}) and various inclinations $I$. Adapted from Ref.~\cite{ripa20}.}
\label{fig:saafluxes}
\end{figure}

A satellite in an LEO can be exposed to intense fluxes of electrons and protons trapped inside the inner Van Allen radiation belt when it passes through the Earth's polar regions or through the SAA (Figure~\ref{fig:SAA1}). As discussed in Section~\ref{sec:leo}, the SAA has been slowly expanding since the discovery of the radiation belts as a consequence of the gradual weakening of the geomagnetic field (Figure~\ref{fig:SAA2}). The particle flux is so intense in the SAA that the scientific instrumentation is often switched off when a high-energy astronomy satellite passes through this region, although some diagnostic data can still be recorded (see, e.g., \cite{abdo09}). 

There are several models to describe the fluxes of trapped particles around the Earth. In Figure~\ref{fig:saafluxes}, we show results of the IRENE (International Radiation Environment Near Earth) AE9/AP9 models \cite{ginet13}, which combine data from 33 satellites to produce realistic probabilities of occurrence for varying flux levels along a user-defined orbit. We see that the orbit-averaged particle fluxes are much higher for polar orbits (inclination $I\sim 90^\circ$) than for equatorial orbits ($I\sim 0^\circ$), by about three orders of magnitude for electrons and two orders of magnitude for protons. The simulated fluxes also increase with altitude, by a factor of 2 to 3 every 50~km \cite{cumani19,ripa20}. 

The radiation environment in an LEO passing at the edge of the SAA was measured by the Particle Monitor (PM) experiment on-board the \textit{BeppoSAX} X-ray satellite almost uninterruptedly between 1996 and 2002 \cite{campana14}. \textit{BeppoSAX} was initially placed in a orbit of about 600~km altitude and $3.9^\circ$ inclination, but the spacecraft altitude significantly decreased with time due to the atmospheric drag, down to about 470~km when the scientific payload was switched off. Measurements of the particle flux as a function of altitude showed a dramatic decrease in intensity, by about an order of magnitude from 600~km to 550~km, which is much more than predicted by the AE9/AP9 models (see Figure~\ref{fig:saafluxes}). The APE8/AP8 models developed at NASA since the 1970's\footnote{See \url{https://ccmc.gsfc.nasa.gov/models/modelinfo.php?model=AE-8/AP-8\%20RADBELT}} seem to be in better agreement with these observations (see \cite{ripa20}). The \textit{BeppoSAX}/PM measurements also showed a clear anti-correlation of the SAA particle flux and the solar activity, as quantified by the mean number of sunspots. These measurements further showed a significant drift westward of the longitude of the SAA maximum by about $0.40^\circ$~yr$^{-1}$, in good agreement with the drift measured with the particle monitor of \textit{RXTE} at a latitude of $-23^\circ$ \cite{furst09}.

Trapped electrons in the inner radiation belt have kinetic energies limited to a few MeV (see Figure~\ref{fig:saafluxes}). These particles can produce significant hard X-ray and soft gamma-ray  background by bremsstrahlung in the spacecraft. They can also contribute to an increase in the leakage current of detectors during SAA transits (see, e.g., \cite{dilillo22}), thus potentially deteriorating the response of high-energy instruments. Trapped protons have energies up to about 1~GeV. These particles can penetrate deep into satellites (the range of a 1-GeV proton in Al is $\sim 1.5$~m) and potentially cause damage to on-board electronics and sensors. In addition, energetic protons in the SAA can build up an instrumental background from the activation of spacecraft materials (see, e.g., \cite{wik14}), as further discussed in the next section. 

\subsection{Delayed background from activation of satellite materials}
\label{sec:activation}

As a spacecraft is bombarded by energetic hadronic particles, nuclear reactions can produce radioactive isotopes in satellite materials, whose decay radiation can be difficult to distinguish from source gamma rays in the MeV range. Since radioactive decay emissions are delayed compared to the interactions of energetic particles with the spacecraft, most of the background from material activation cannot be suppressed by an anti-coincidence detector of charged particles. The gamma and $\beta^+$ radioactivities can induce several hits in coincidence in the gamma-ray detectors, which can be confused with Compton interactions from celestial gamma-ray photons. The decay of radioactive nuclei in the detectors by the emission of $\alpha$ or $\beta^-$ particles, or by electron capture, without the concomitant emission of a gamma-ray photon, produces single events that can be rejected from the background of a Compton telescope, but not easily from the one of a coded-mask instrument. A strong background line is generally expected to be the $511$~keV positron-electron annihilation line resulting from the decay of various $\beta^+$ radioisotopes (e.g. \cite{weidenspointner03,cumani19}). 

\begin{figure}[t]
\centering
\includegraphics[width=0.6\textwidth]{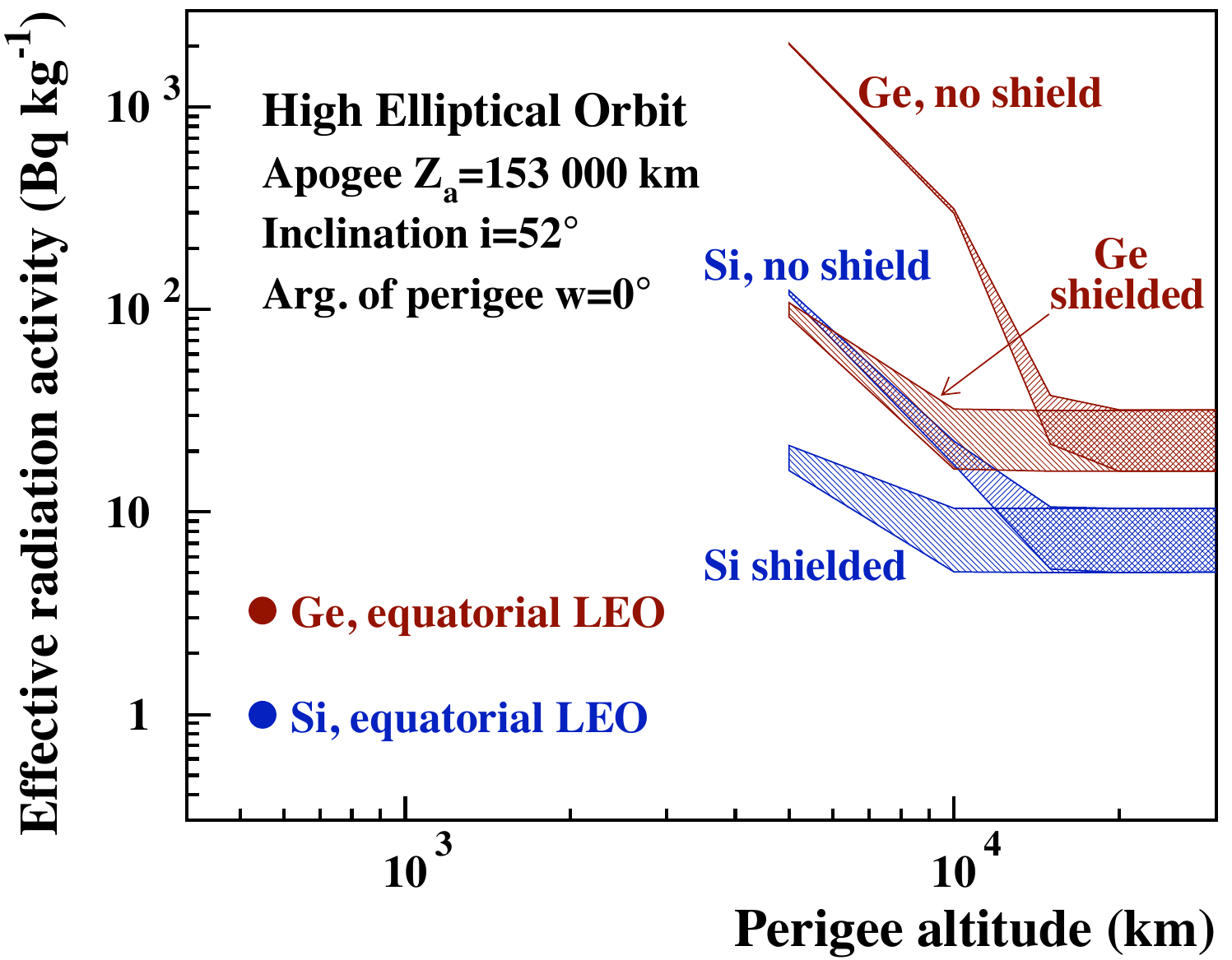}
\caption{Effective radiation activity of Si and Ge after $1$~yr of proton irradiation as a function of the perigee altitude of a high elliptical orbit like that of the \textit{INTEGRAL} satellite (see Sect.~\ref{sec:ellipticalorbits}). Two irradiation conditions are considered: either the semiconductors are directly exposed to the proton flux or they are shielded from the radiation environment by a C layer of $1.3$~g~cm$^{-2}$ (see text). The hatched areas reflect the uncertainties arising from the solar activity. Also shown is the activity of these materials in an equatorial LEO at an altitude of 550 km. In this case, the results with and without the C shielding are almost identical.}
\label{fig:activation_zp}
\end{figure}

The importance of this delayed background depends strongly on the satellite orbit and the nature of the detectors. To illustrate these points, we show in Figure~\ref{fig:activation_zp} calculated activation of Si and Ge (two common detector materials) after 1 year of proton irradiation in various orbits. Two cases were considered for the irradiation setup: either that a representative block of $100$~kg of Si or Ge is directly exposed to the radiation environment, or that the detectors are shielded from the proton flux by a C layer of $1.3$~g~cm$^{-2}$. The latter is representative of an anti-coincidence system made of about $1$-cm thick plastic scintillator panels covering the gamma-ray instrument. The radioisotope production is calculated by taking into account both the nuclear reactions induced by the bombarding protons and the reactions from the secondary protons and neutrons produced in the detectors. The cross sections of the proton- and neutron-induced spallation reactions were obtained from the TALYS nuclear reaction code\footnote{\url{http://www.talys.eu/}} below 250 MeV and from the Li\`ege intranuclear cascade code\footnote{\url{https://irfu.cea.fr/dphn/Spallation/incl.html}} at higher energies. Once the production of the radioisotopes was evaluated, the decay radiation from these nuclei, as well as from their daughter isotopes, was computed using the NuDat library\footnote{\url{https://www.nndc.bnl.gov/nudat3/}}. The ``effective radiation activity" shown in Figure~\ref{fig:activation_zp} is defined as the rate of radioactive events that can produce two hits or more in the detectors and thus be confused with Compton events from cosmic gamma rays (J. Kiener, private communication). 

Two kinds of orbits have been considered for Figure~\ref{fig:activation_zp}: an equatorial LEO at an altitude of $550$~km, and a high elliptical (apogee altitude $Z_a=1.53\times 10^5$~km), high inclination ($I=52^\circ$) orbit like that of the \textit{INTEGRAL} satellite (see Sect.~\ref{sec:ellipticalorbits}). \textit{INTEGRAL} was launched with an initial perigee altitude $Z_p \simeq 9~000$~km and initially spent most of the time (more than $80$\%) above an altitude of $60~000$~km, well outside the Earth's radiation belts. But the perigee altitude varied considerably throughout the mission as a result of the Earth's oblateness and the luni-solar gravitational perturbations, from $Z_p \sim 2~000$ to $13~000$~km. We see in Figure~\ref{fig:activation_zp} that when the detectors are shielded from the low-energy proton flux~--~the assumed C layer of $1.3$~g~cm$^{-2}$ stops protons of less than about $35$~MeV~--~the passage through the Earth's radiation belts has no significant effect on the detector activation as long as $Z_p > 10~000$~km. However, we also see in Figure~\ref{fig:activation_zp} that the predicted effective radiation activities outside the belts are about 5 to 10 times higher than those calculated for detectors on an equatorial LEO, depending on the solar activity. 

Figure~\ref{fig:activation_zp} also shows that the activation of Ge is higher than that of Si, by a factor of about $3.1$ for high-Earth orbit (see the results for $Z_p = 30~000$~km), $3.3$ for an equatorial LEO, and reaching about $17$ for an \textit{INTEGRAL}-like high elliptical orbit with $Z_p = 5~000$~km and no shielding of the detectors. The main species contributing to the Si activity are $^{30}$P (half-life $T_{1/2}=2.498$~min), $^{29}$P ($4.142$~s), $^{28}$Al ($2.245$~min) and $^{27}$Si ($4.15$~s). Those for the Ge activity are $^{74}$As ($17.77$~d), $^{72}$As ($26.0$~h), $^{73}$Ge$^{\rm m}$ ($0.499$~s), $^{73}$As ($80.30$~d), $^{76}$As ($1.0942$~d) and $^{70}$As ($52.6$~m). It is remarkable that the radioisotopes produced in Si have much shorter lifetimes than the main radioisotopes produced in Ge. As a result, the activity of Si detectors on LEOs passing through the South Atlantic Anomaly (SAA; Sect.~\ref{sec:saa}) is expected to decrease rapidly when the satellite leaves the SAA, whereas the Ge activity should accumulate over much longer time periods. 

\begin{figure}[ht]
\centering
\includegraphics[width=0.88\textwidth]{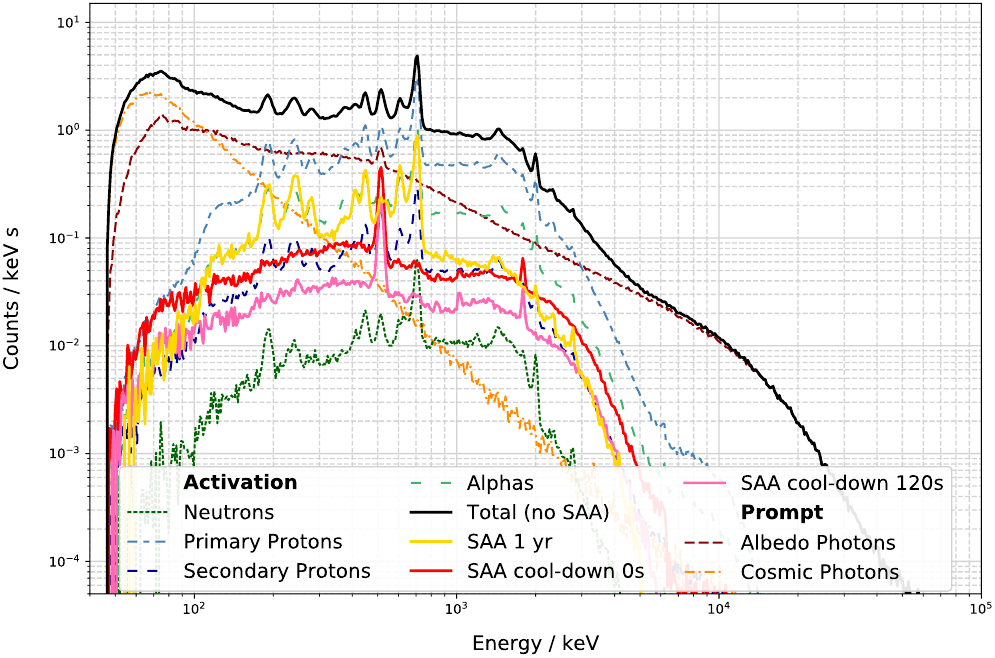}
\includegraphics[width=0.88\textwidth]{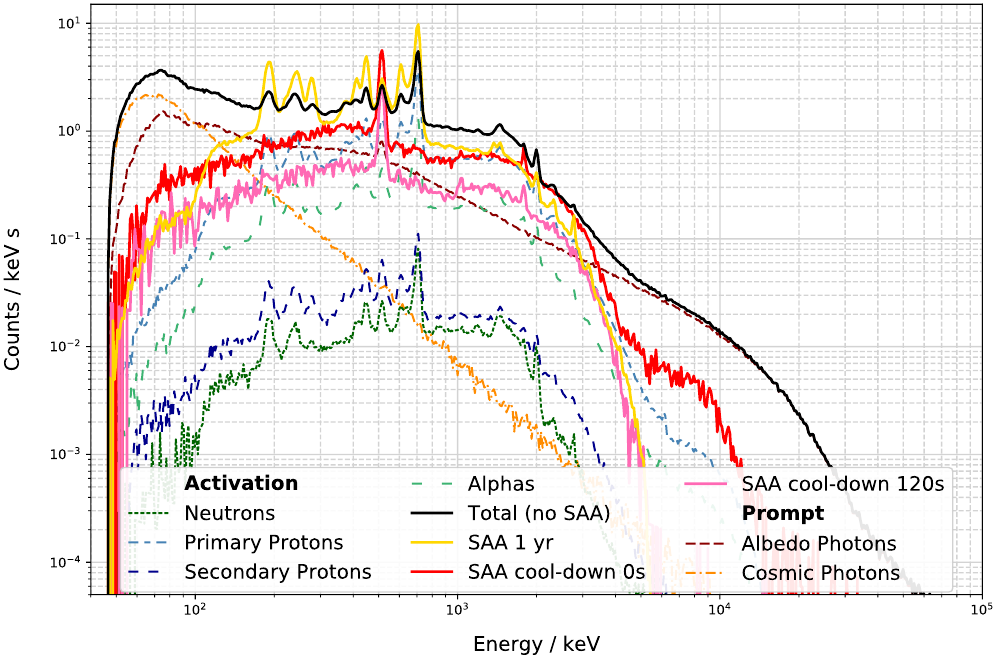}
\caption{Spectra of the reconstructed background Compton events in the e-ASTROGAM gamma-ray telescope for an LEO of 550 km altitude and inclination of $0^\circ$ (\textit{top}) and $15^\circ$ (\textit{bottom}). The background from the spacecraft activation is calculated from the fluxes of primary protons and alpha-particles (i.e. Galactic cosmic rays), secondary protons and neutrons, as well as protons trapped in the SAA. The prompt background is from the extragalactic gamma-ray emission and the Earth's albedo photons. The ``total" emission is obtained from the sum of all the contributions except that from activation in the SAA. The SAA contribution is divided into short-term, with $0$~s or $120$~s cool down time after the SAA passage, and long-term background (SAA 1 yr). Figure adapted from Ref.~\cite{cumani19}.}
\label{fig:activation_ea}
\end{figure}

In Figure~\ref{fig:activation_ea}, we show simulated background spectra of the e-ASTROGAM telescope \cite{deangelis17} for two LEO with different inclinations, $I=0^\circ$ and $15^\circ$. e-ASTROGAM is a mission proposal for gamma-ray observations in the energy range between 300~keV and 3~GeV, with a gamma-ray instrument made of three main detectors: a tracker composed of 56 Si layers supported by a mechanical structure in polymer resin and C fibers, a calorimeter comprising a large number of thallium activated cesium iodide crystals (CsI(Tl)) read out by Si drift detectors, and an anti-coincidence detector made by plastic scintillators coupled to Si photomultipliers covering both the top and the sides of the instrument. The simulations were performed with the Medium Energy Gamma-ray Astronomy library (MEGAlib) toolkit \cite{zoglauer06}. For the activation of the spacecraft caused by protons trapped in the SAA, two types of background were considered: a short-term background due to the production of short-lived radioisotopes, which decreases rapidly after the satellite exits the SAA, and a long-term background, which gradually accumulates during the mission. For the latter, the radioisotope production was calculated assuming an irradiation of the spacecraft by the SAA protons for $72$~days, which corresponds to the approximate total time the satellite spends in the SAA during one year in an equatorial LEO ($\sim 19$~min per orbit for a $550$~km orbit of period $96$~min). This irradiation period was followed by a cool down time of 48~min in the simulation (half an orbital period), in order to consider only the decay of relatively long-lived radioisotopes \cite{cumani19}.   

We see in Figure~\ref{fig:activation_ea} that the Earth albedo and extragalactic gamma-rays are predicted to be the main components of the background below $\sim 150$~keV and above $\sim 4$~MeV, but activation should be dominant between these two energies. The two most prominent lines that appear in the activation spectra are at $511$~keV and $700$~keV. The $511$~keV line is produced following the $\beta^+$ decay of radioisotopes such as $^{11}$C ($T_{1/2}=20.364$~min) and $^{15}$O ($122.24$~s). The 700 keV line, which is prominent in all activation spectra except those for the short-term SAA background, is mainly produced by the electron-capture decay of two long-lived isotopes abundantly created in the CsI(Tl) crystals of the calorimeter: $^{132}$Cs ($6.480$~d) and $^{126}$I ($12.93$~d). The first radioisotope produces nuclear gamma rays of $668$~keV at the same time as K$\alpha$ X-rays of $30$~keV. The second one produces $666$~keV gamma rays together with $27$~keV X-rays.

For an equatorial LEO (\textit{upper panel}), the main contributor to the total activation comes from the decay of radioisotopes produced by primary Galactic cosmic-ray protons. The total rate of events from the long-term SAA activation is about a fifth of that due to cosmic-ray protons. The total count rate of the short-term SAA background is lower than that of the long-term SAA background by a factor of at least $1.5$, even without considering any cool down time after the SAA passage. On the other hand, for an inclination of $15^\circ$ (\textit{lower panel}), the long-term SAA activation is predicted to be higher than the sum of all the other components in the $\sim 150$~keV to $\sim 720$~keV energy range. Moreover, the short-term SAA activation is expected to make a major contribution to the $511$~keV line emission for several minutes after the passage through the SAA. In terms of background, an LEO with a low inclination ($I < 5^\circ$; see \cite{cumani19}) should generally be the best orbit for a MeV gamma-ray mission.

\section{Conclusions}
\label{sec:conclusions}

In this chapter we have presented the different orbits available for gamma-ray space missions, and described in detail the properties of the background particles and radiation that limit the detection sensitivity of space-borne gamma-ray instruments. We also discussed the value of stratospheric balloon experiments to prepare and sometimes supplement space missions. A good knowledge of the environment of high-energy experiments is all the more necessary as highly energetic particles and gamma-rays can also damage detectors and components of electronic integrated circuits.

In terms of background count rate, the best environmental conditions of a gamma-ray mission are found in near-equatorial, low altitude (typical altitude of $500$--$600$~km) circular orbit. A satellite on such LEO is well shielded from low-energy cosmic rays and solar energetic particles by the Earth's magnetic field. Proton and alpha-particle fluxes beyond the Earth's radiation belts are typically an order of magnitude higher than those in equatorial LEO, resulting in lower activation of spacecraft materials and  lower instrumental background in low- than in high-Earth orbits, especially in the MeV gamma-ray range. Anyway, this conclusion is valid only for LEO inclinations below $\sim 10^\circ$, as for higher inclinations the orbit-averaged flux of protons trapped in the inner radiation belt can exceed that in high-Earth orbit. The intense electron and proton fluxes experienced by a gamma-ray satellite crossing the South Atlantic Anomaly also have the drawback of significantly limiting the observation duty cycle. 

A satellite in high-Earth orbits or at the L1 or L2 Lagrange points can benefit from a more stable environment, in particular not subject to the day-night variations of the equatorial LEO and the accompanying temperature changes. Moreover, in these distant orbits the lack of Earth's albedo radiation and secondary particles produced in the atmosphere can allow the conception of gamma-ray instruments with very large fields of view capable of continuously monitoring almost the entire gamma-ray sky with good detection efficiency. But a caveat is due to direct exposure to solar energetic particles that can be a source of an additional background component 
not shielded by the Earth's magnetic field. During periods of intense solar activity this may limit the observing time, as happened for example for \textit{INTEGRAL} observations. In view of the above mentioned issues, the choice of the ``best orbit" for a gamma-ray satellite has to be optimised taking into account not only the 
detection sensitivity
of the on-board experiments, but also other important requirements such as:
\begin{itemize}
    \item The required duration of uninterrupted observations: LEOs have an average science window for observations of less than 1~hour, while high elliptical orbits can provide observations lasting up to 3~days and L1--L2 orbits can allow even longer uninterrupted observations. 
    \item The required sky coverage: LEOs are compatible with a coverage of about half of the sky at any moment, while the field of view in high elliptical orbits is barely obscured by the Earth when the satellite is transiting with high speed close to perigee, granting large part of the orbit with almost all-sky visibility. Thus, \textit{Fermi}/GBM monitors the whole sky with about 60\% duty cycle, while it is 85-90\% for the anti-coincidence shield of \textit{INTEGRAL}/SPI. 
    \item The main operational constraints, i.e. satellite lifetime, tracking station availability and hand-over constraints, data download requirements (i.e. real-time versus in-board storage) and telecommand need.
\end{itemize}
In the era of multi-messenger astronomy recently opened by the simultaneous detection of gamma-rays and gravitational waves, as well as the detection of high-energy neutrinos, 
it is essential to have in-space instruments with good imaging and monitoring capabilities able to detect and localise 24~h/7~d weak signals from energetic transients in the energy range from $\sim 0.1$~MeV to a few hundred MeV. 
The required highly efficient coverage of the whole sky should be associated with excellent, omni-directional sensitivity above 100~keV and rapid reaction time for Target-of-Opportunity observations.



\section{Cross-References}

\begin{itemize}
    \item Volume 1, Section II:
    \begin{itemize}
        \item Chapter: In-orbit background
    \end{itemize}
    \item Volume 2~--~Chapter: History of Gamma-ray Astrophysics
    \item Volume 2, Section IV:
    \begin{itemize}
        \item Chapter: Telescope concepts in gamma-ray astronomy
        \item Chapter: Compton telescopes 
        \item Chapter: Pair-creation telescopes
        \item Chapter: Gamma-ray detector and mission design simulations
    \end{itemize}
    \item Volume 2, Section V:
    \begin{itemize}
        \item Chapter: The COMPTEL instrument on the CGRO mission
        \item Chapter: The INTErnational Gamma-Ray Astrophysics Laboratory (INTEGRAL)
        \item Chapter: The AGILE mission
        \item Chapter: The Fermi Gamma-ray Space Telescope mission
    \end{itemize}
    \item Volume 3a, Section IX:
    \begin{itemize}
        \item Chapter: Galactic cosmic rays
    \end{itemize}
    \item Volume 3b, Section XV:
    \begin{itemize}
        \item Chapter: Surveys of the Cosmic X-ray Background
        \item Chapter: Multimessenger observations
    \end{itemize}
\end{itemize}

\begin{acknowledgement}
The research presented in this chapter has received funding from the European Union’s Horizon 2020 Programme under the AHEAD2020 project (grant agreement n. 871158). 
PU acknowledge the continuous support of the ASI-INAF Agreement N.2019-35-HH.0.
We thank Pierre Cristofari for his careful reading of the manuscript. 
\end{acknowledgement}

\bibliography{references}
\end{document}